# From Aircraft Tracking Data to Network Delay Model: A Data-Driven Approach Considering En-Route Congestion

Yu Lin[1], Lishuai Li[2*], Pan Ren[3], Yanjun Wang[4], W.Y. Szeto[5]

*Abstract:*

En-route congestion causes delays in air traffic networks and will become more prominent as air traffic demand will continue to increase yet airspace volume cannot grow. However, most existing studies on flight delay modeling do not consider en-route congestion explicitly. In this study, we propose a new flight delay model, Multi-layer Air Traffic Network Delay (MATND) model, to capture the impact of en-route congestion on flight delays over an air traffic network. This model is developed by a data-driven approach, taking aircraft tracking data and flight schedules as inputs to characterize a national air traffic network, as well as a system-level model approach, modeling the delay process based on queueing theory. The two approaches combined make the network delay model a close representation of reality and easy-to-implement for what-if scenario analysis. The proposed MATND model includes 1) a data-driven method to learn a network composed of airports, en-route congestion points, and air corridors from aircraft tracking data, 2) a stochastic and dynamic queuing network model to calculate flight delays and track their propagation at both airports and in en-route congestion areas, in which the delays are computed via a space-time decomposition method. Using one month of historical aircraft tracking data over China's air traffic network, MATND is tested and shows to give an accurate quantification of delays of the national air traffic network. "What-if" scenario analyses are conducted to demonstrate how the proposed model can be used for the evaluation of air traffic network improvement strategies, where the manipulation of reality at such a scale is impossible. Results show that MATND is computationally efficient, well suited for evaluating the impact of policy alternatives on system-wide delay at a macroscopic level.

## 1 Introduction

Air travel demand has been increasing steadily over the past ten years. Such rapid growth resulted in greater degrees of congestion in air traffic networks; subsequent flight delay wastes travelers' time and fuel and takes a toll on economic activities and the environment. To reduce flight delays, it is critical to analyze the capacity, efficiency, and congestion risks of the existing air traffic network, and then carry out strategic planning and tactical management accordingly.

[1] PhD student, School of Data Science, City University of Hong Kong, Hong Kong SAR. Email: yulin8-c@my.cityu.edu.hk

[2] Assistant Professor, School of Data Science, City University of Hong Kong, Hong Kong SAR. (sponsor)
Assistant Professor, Faculty of Aerospace Engineering, TU Delft, Delft, Netherlands. (current affiliation) Email: lishuai.li@tudelft.nl
* Corresponding Author

[3] Department of Systems Engineering and Engineering Management, City University of Hong Kong, Hong Kong SAR. Department of Operations and Planning, SF Express Co., Ltd, China. Email: panren@sfmail.sf-express.com

[4] Associate Professor, College of Civil Aviation, Nanjing University of Aeronautics and Astronautics, China. Email: ywang@nuaa.edu.cn

[5] Professor, Department of Civil Engineering, The University of Hong Kong. Hong Kong SAR. Email: ceszeto@hku.hk

Among many factors contributing to flight delays, en-route congestion becomes prominent in crowded airspace in the world, including Europe, the US, and China. Airspace may seem plentiful compared to the amount of aircraft maneuvering in it. However, airspace does get congested in areas with a high flight density, such as the merging areas of air traffic flows from different directions. Fig. 1 illustrates how en-route congestion is formed. In this figure, Sector A reaches its capacity (i.e., 15 aircraft). No other aircraft is allowed to enter Sector A. Aircraft in Sectors B and C that need to pass Sector A now have to divert or hold. In practice, this problem is solved by adjusting the departure times and routes of the involved flights to balance the air traffic flow among different sectors, which may result in en-route delays. For instance, when a sector is predicted to undergo congestion, the involved flights are delayed at the departure airports or their routes are partly changed to avoid aggravating the congestion of this sector.

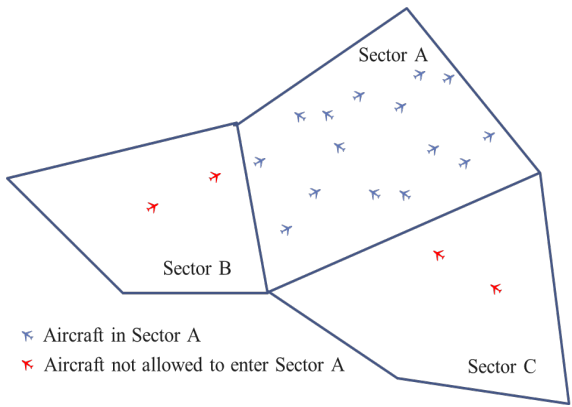

Fig. 1 Illustration of en-route congestion: Sector A reaches its capacity. No aircraft is allowed to enter Sector A at the moment.

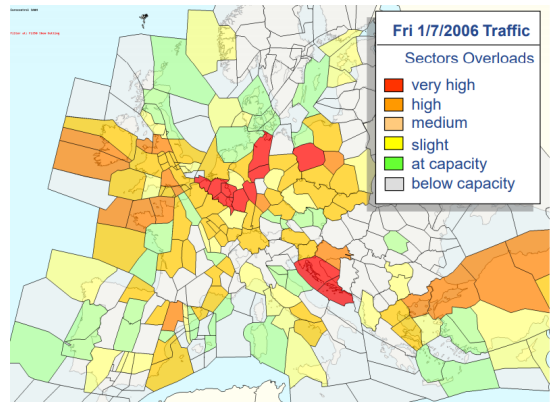

Fig. 2 Airspace overload in Europe (Russo, 2016).

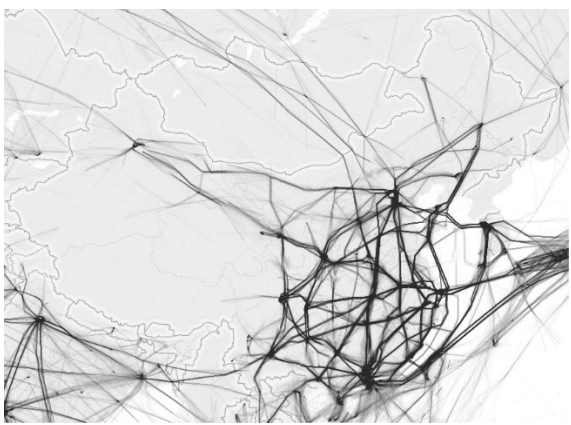

a) China

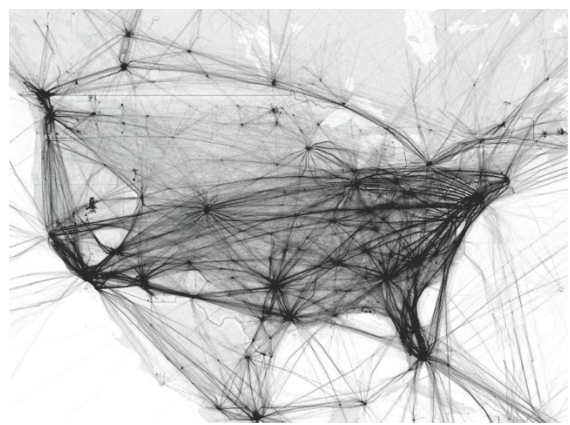

b) USA

Fig. 3 Flight track visualization using one-month of actual flight tracks (Ren and Li, 2018).

According to a report by CAPA, an aviation consultancy, "The biggest contributor to aviation delay in Europe is a lack of Air Traffic Control (ATC) capacity. En-route ATC capacity accounted for 25.5% delays in 2017, and airport capacity another 15.5% of interruptions." (CAPA, 2018) Fig. 2 shows the airspace overload situation in Europe. The en-route delay has also been reported as the fastest-growing source of delays in Europe, with an average yearly rate of 17% from 2005 to 2010 (Eurocontrol, 2010). Similar developments have also been seen in some areas of the US. This is the main reason why both the US and Europe have initiated huge programs called Next Generation Air Traffic Control (NextGen) in the US and Single European Sky ATM Research (SESAR) in Europe (FAA, 2019; SESAR Joint Undertaking, 2020). One of the common aims shared in both programs is to increase airspace capacity. In China's case, the airspace available for airline flights is even more limited - about 80% of airspace is controlled by the military, not open for civilian traffic (Hsu, 2014). There is a shortfall in airspace and air routes for civilian traffic in a previous study using actual flight tracking data (Ren and Li, 2018). As shown in Fig. 3, the en-route airspace resources are much more limited in China than in the USA, resulting in a higher risk of en-route congestion. The Chinese government has pledged to integrate the civil and military management of airspace, and to release more airspace for commercial use. However, limited methods are available to model the delays of a large air traffic network with constrained airspace capacity. The focus of this study is to model delays and their propagation over a national air traffic network where en-route resources face capacity shortfalls. The model can be used to evaluate different capacity improvement strategies, e.g., adding air corridors, increasing sector capacities, and shifting towards dynamic airspace configuration, from the perspective of delay reduction.

Air Traffic Management (ATM) studies related to flight delay are normally framed as optimization problems, which try to solve the mismatch between supply (airspace and airport capacity) and demand (scheduled and non-scheduled flights). These studies can be broadly categorized into two levels: strategic ATM and tactic ATM. The majority belongs to tactic ATM, which seeks to minimize delays (or delay-related costs) via Air Traffic Flow Management (ATFM) interventions, e.g., aircraft re-routing, speed adjustment, ground holding, runway and terminal area scheduling and sequencing, or through airline-initiated actions, e.g., flight swaps and cancellation, given a capacity and demand scenario. Integer programming models are built to optimize flight movement via ground and airborne holding, speed control, and re-routing (Bertsimas et al., 2011, 2008; Bertsimas and Patterson, 1998; Lulli and Odoni, 2007). In a later model, Marla et al. (2017) focused on speed changes and flight departure holding from the perspective of airlines.

On the strategic ATM level, there is a limited number of studies in the literature. Bolić et al. (2017) presented an integer programming model for strategic redistribution of demand (flight schedule shifts and route choice control) under hard capacity constraint (nominal declared capacity of airports and sectors) considering the whole network. Starita et al. (2020)'s approach considered the strategic planning on both supply and demand sides, minimizing the sum of capacity provision cost (that reflects the sector-hours budgeted for each airspace for a specific day in the future) and displacement cost (including both additional fuel cost and delay cost due to rerouting and delay) for a network. However, a key element in the strategic ATM optimization models, the delay modeling on a large network, was not the focus of these studies. Neither studies considered stochastic queue dynamics at all possible bottlenecks, e.g. airport, terminal airspace, en-route airspace, and delay propagation over the network, which has significant impacts on optimal policies. We have found only one study on strategic ATM optimization that considers stochastic queue dynamics so far. Jacquillat and Odoni (2015) developed an approach that jointly optimizes the airport's flight schedule at the strategic level and the utilization of airport capacity at the tactical level to mitigate delays. However, the delays caused by the airspace (en-route and terminal) capacity constraint are not included in this work.

Focusing on the flight delay modeling, rather than ATM optimization, there are three streams of research. The first stream involves agent-based simulations. The state-of-the-art models in this category are the two agent-based models that include details of the entire National Airspace System (NAS), the Airspace Concept Evaluation System (ACES) (Meyn et al., 2006) and the Future ATM Concepts Evaluation Tool (FACET) (Bilimoria et al., 2001), both of which are currently used by the National Aeronautics and Space Administration (NASA) and the Federal Aviation Administration (FAA). ESCAPE is a scalable EUROCONTROL air traffic management (ATM) real-time simulation platform (Eurocontrol, 2020). More recently, Fleurquin and Campanelli's team developed two agent-based models to simulate flight delay propagations in the US and European networks (Campanelli et al., 2016; Fleurquin et al., 2014, 2013). These simulation models enable a wide range of capabilities and provide accurate results with detailed information. However, because of the extensive input preparations required, these simulation tools are cumbersome and cannot be used for rapid evaluations of policy-oriented ATM improvement strategies.

The second stream uses statistical, data mining, or econometric methods and historical data to model and predict delays. Xu et al. (2008) used regression models to estimate local delays and propagated delays at major airports in the USA. Rebollo and Balakrishnan (2014) used a random forest approach to predict departure delays from 2–24 h into the future. Some studies have focused on scheduling from the perspective of airlines. Beatty et al. (1999) and AhmadBeygi et al. (2008) used delay trees to track how delays propagate within an airline's network. Fricke and Schultz (2009) analyzed the individual inbound delay's impact on the turnaround process duration and stability. Kafle and Zou (2016) proposed an econometric model to analyze delay propagations. Results provide estimates on how much propagated delay will be generated out of each minute of newly formed delay, for the US domestic aviation system as well as for individual major airports and airlines. Unfortunately, these methods are not suited for evaluating improvement strategies in the ATM procedure and infrastructure, because the ATM procedure and infrastructure are not explicitly modeled in these methods.

The third stream develops analytical models based on queuing networks, which is the most promising approach for addressing the problems of interest in this study. The basic idea of these models is to treat airports in the network as a set of individual interconnected queuing systems, and recursively compute the delay propagation over the entire network considering the dynamic characteristics of airport capacity and aircraft queues. Peterson et al. (1995a, 1995b) developed a recursive approach based on a semi-Markov model to compute the queue lengths and waiting times of aircraft landings in a hub-and-spoke configuration with a single hub airport. A national-scale airport network model, LMINET, was developed by Long et al. (1999) and an updated version was described by Long and Hasan (2009). Tandale et al. (2008) presented a family of queuing network models for incorporating trajectory uncertainties at national, regional, and local scales. Pyrgiotis et al. (2013) developed the Approximate Network Delays (AND) model, a stochastic and dynamic queuing network model to compute local and propagated delays within the NAS using aircraft itinerary information. This model was first conceptualized by Malone (1995) in her Ph.D. thesis. The AND model is very fast in computation, thus making possible the exploration at a macroscopic level of the impact of a large number of scenarios and policy alternatives on system-wide delays. However, these models are generally based on airport networks and do not take en-route constraint into account. They are inadequate in cases where the airspace is restricted and en-route congestion is significant. Following this research approach, this study models delays and their propagation over a national air traffic network as a set of individual interconnected queuing systems, focusing on the impact of en-route capacity shortfalls.

Incorporating en-route constraint in delay modeling presents significant challenges. If we model the capacity and demand at each sector's level, a significant amount of detailed information is needed from the air navigation service provider (ANSP) to build up the en-route structure, such as sector maps, published airways, navigational aids, metering fixes, and individual flight trajectories. In addition, the capacity of each sector is difficult to estimate. Majumdar and Polak (2001) provided a framework for modeling Europe's airspace en-route capacity by considering the factors that affect controller workload and then using a model of controller workload, aided by the appropriate analytical techniques, to estimate airspace capacity. Cho et al. (2011) incorporated the convective weather effects into an analytical sector workload model to estimate en-route capacity under convective weather. Welch (2015) described a new workload-based capacity model that improves upon the FAA's current Monitor Alert capacity model.

Unlike the existing papers, this paper proposes a new data-driven approach to learn the en-route structure from large-scale air traffic surveillance data and capture the en-route delay effects through conceptual en-route "congestion points", rather than modeling the capacity and demand of each sector. Our proposed approach is enabled by the availability of aircraft tracking data. While FAA, Eurocontrol, and other ANSPs normally maintain the most comprehensive and highest-quality air traffic surveillance data for a particular region, it is increasingly easy to obtain air traffic surveillance data via public and commercial sources. Several commercial and non-commercial websites track aircraft positions via the crowd-sourced distributed networks of Automatic Dependent Surveillance - Broadcast (ADS-B) receivers and make the tracking data publicly available, e.g., FlightAware, Flightradar24, OpenSky Network, and VariFlight.

There are a few studies that used data-driven approaches to learn typical air traffic flows and patterns, mainly in the terminal area (Conde Rocha Murca et al., 2016; Eckstein, 2009; Enriquez, 2013; Gariel et al., 2011; Rehm, 2010; Salaun et al., 2012). However, there have been no attempts to connect these flow identification tools with network delay modeling. Andrienko et al. (2017) and Buschmann et al. (2016) developed visualization tools to filter and perform cluster analysis on spatial-temporal trajectory data for air-traffic analysis. Gariel et al. (2011) and Salaun et al. (2012) focused on data-driven flow modeling to support air traffic controller monitoring and managing airspace. Murça et al. (2018a) and Murça et al. (2018b) proposed a data mining framework to characterize air traffic flows using aircraft tracking data, where a density-based clustering algorithm is used to identify major flight trajectory patterns in the airspace of multi-airport systems.

On the topic of typology identification on trajectory data, relevant papers are scattered in the field of computer science, transportation research, and geographic information systems. Computer science papers focus on the development of clustering algorithms to summarize movement data (Andrienko et al., 2017; Ferreira et al., 2013; Palma et al., 2008; Pelekis et al., 2009). The key challenges lie in scalability for processing large datasets, effective measures of similarity between trajectories, and semantic information extraction. Studies in the field of transportation research tackle travel pattern characterization (Kim and Mahmassani, 2015), route identification (Chen et al., 2011), and map construction (Buchin et al., 2017) from vehicle trajectories using clustering and other data mining techniques. However, no study has been conducted to identify traffic flows and link them to network typology (nodes and links) from trajectory data, to the best of our knowledge.

Therefore, to fill in the research gap, we propose a novel flight delay model, named as Multi-layer Air Traffic Network Delay (MATND) model, to capture the impact of en-route congestion on flight delays over an air traffic network and to automatically learn the structure of a national air traffic network from historical aircraft tracking data via unsupervised learning methods. The model includes two parts: 1) a data-driven method to learn and construct a national air traffic network consisting of airports and en-route congestion points as nodes and operational air routes as links. The operational air routes are extracted using clustering algorithms and then a score function is constructed to identify conceptual en-route congestion points from historical aircraft tracking data; and 2) a stochastic and dynamic queuing network model to compute delays and predict their propagation over the constructed network. The queuing network model estimates delays at each airport and en-route congestion point and tracks the propagation of these delays and their impacts on subsequent flight operations. To demonstrate its performance, the model is implemented on an air transport system consisting of the 56 busiest airports in China (as ranked by annual passenger traffic (Civil Aviation Administration of China (CAAC), 2018, 2016). Using the proposed MATND model, decision-makers can predict the risk of flight delays under various operational scenarios, such as network structure modifications, infrastructure improvements, or changes in air traffic management procedures.

This work contributes to the literature of flight delay modeling in the following ways: 1) it is the first attempt to capture en-route congestion in flight delay modeling; 2) an unsupervised learning scheme is proposed to extract the nodes (airports, en-route congestion points) and links (operational air routes) of a national air traffic network from aircraft tracking data; 3) analysis based on real-world data provides insights on current bottlenecks of the air traffic network and potential improvement strategies in China.

The remainder of this paper is organized as follows. Section 2 describes the methods developed to model delays and their propagation over a national air traffic network considering en-route constraint using historical aircraft tracking data and flight schedule data. Section 3 shows the superior of the proposed MATND model to the AND model via a theoretical proof on a simplified network. Section 4 demonstrates the importance of including en-route congestion in a network delay model using a stylized network. The implementation and validation of the proposed model for the air transport system in China are discussed in Section 5. Moreover, three scenario analyses are conducted to show how the proposed model is used to evaluate the impact of infrastructure improvements and network structure modifications in Section 6. Finally, Section 7 summarizes our study and suggests future research directions.

## 2 Methodology

This section describes the methods developed to model delays and their propagation over a national air traffic network considering en-route constraint using the historical aircraft tracking data and flight schedule data. The framework of the proposed model, Multi-layer Air Traffic Network Delay (MATND) model, is shown in Fig. 4. The model includes two parts: 1) a data-driven method to construct a national multi-layer air traffic network consisting of airports and en-route congestion points as nodes, and operational air routes as links; 2) a stochastic and dynamic queuing network model to simulate delays and their propagation over the network. These parts are explained in detail in the following subsections.

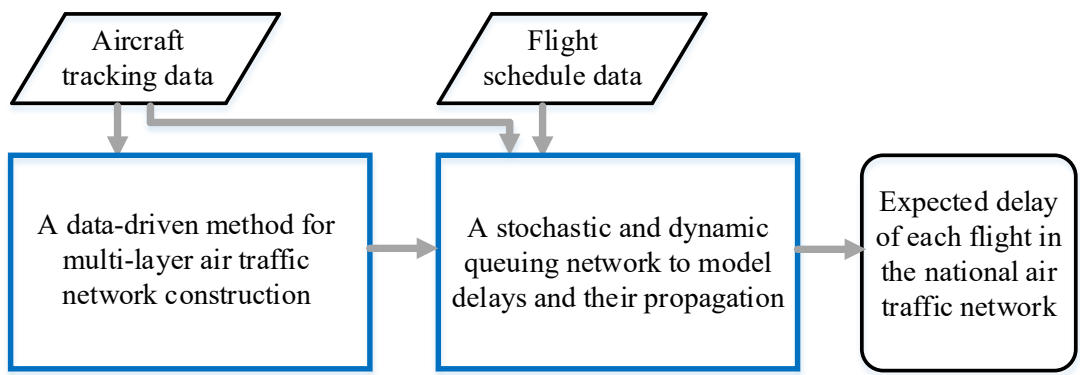

Fig. 4 Framework of the Multi-layer Air Traffic Network Delay (MATND) model.

### 2.1 Input dataset

Only datasets available to the public are used in this method. The details of the input dataset used in this study are described below:

**(1) Aircraft tracking data**

The aircraft tracking data used in this study were collected from Flightradar24 every minute for 30 consecutive days (November 1-30, 2016), including flight ID, latitude, longitude, altitude, speed, origin airport, destination airport, aircraft type, and aircraft registration. We selected flights departing from and arriving at the 56 busiest airports in China. The database was used to identify operational air routes and locate the geographic positions of en-route congestion points.

**(2) Flight schedule data**

Flight schedule data over the same spatial and temporal scales as the aircraft tracking data were collected from VariFlight, including flight ID, origin airport, destination airport, scheduled departure/arrival time, and real departure/arrival time for each flight. The database was used to determine the demand and service profiles of each queuing system in the stochastic and dynamic queuing network, and this information was then used to estimate the demand rates and the service rates.

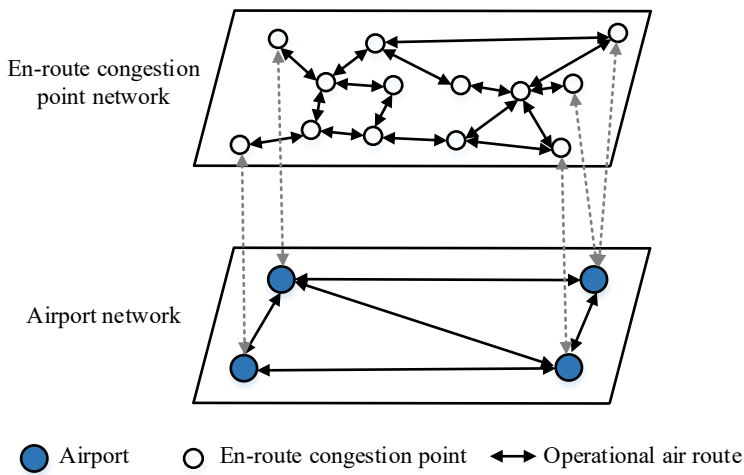

Fig. 5 Multi-layer air traffic network illustration.

**2.2 A data-driven method for multi-layer air traffic network construction**

The first part of the framework involves developing a data-driven method to construct a multi-layer air traffic network of China by considering en-route airspace constraint. The network consists of airports, operational air routes, and en-route congestion points. The multi-layer air traffic network is illustrated as Fig. 5. Instead of just using airports as the nodes of the network, we also incorporate en-route congestion points identified using historical data. Fig. 6 shows the workflow of constructing this multi-layer air traffic network. We first filter trajectories by airport pair and resample the trajectories via interpolation to make the trajectory vectors have equal length. Then, clustering analysis is performed on the resampled flight trajectories to identify operational air routes. After that, the whole air space is divided into different cells where a score function is computed to evaluate the congestion level of each grid. Those cells with a high score are considered as the bottlenecks of the system, and are recognized as en-route congestion points in this method.

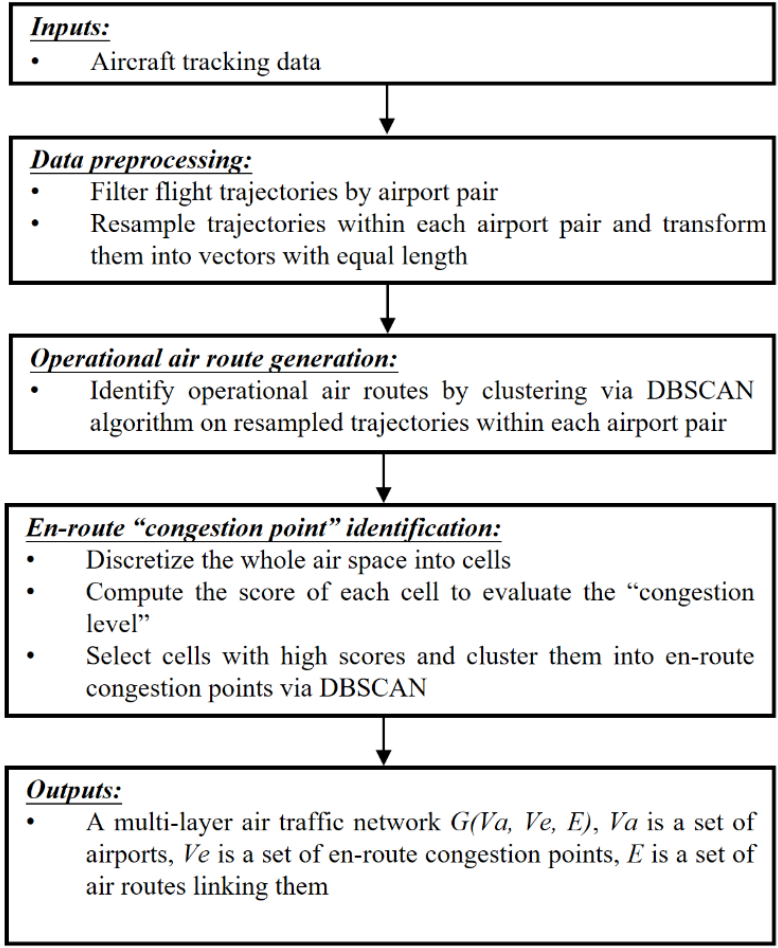

Fig. 6 Multi-layer air traffic network construction.

### 2.2.1 Operational air route identification

We identify the operational air routes used by airline flights by applying a clustering algorithm, DBSCAN (Ester et al., 1996), on the flight trajectory data within each airport pair. No prior information on the airspace structure is needed to identify these operational air routes using this method. Using DBSCAN, the number of air routes linking airport pairs can be automatically determined based on the data distribution patterns. Abnormal flight tracks can emerge as a result of vectoring or other special conditions, and these are treated as outliers by DBSCAN. Previous studies have demonstrated that DBSCAN is an effective technique for identifying operational air routes (Murça et al., 2016; Ren and Li, 2018).

Two key input parameters, *Minpt* and *Epsilon*, may result in changes to the clustering result. *Epsilon* is the maximum radius of neighborhood distance between points and *Minpt* is the minimum number of points in an *Epsilon* neighborhood. For our dataset, the clustering result is not sensitive to *Minpts* when it is between 3 and 10, so we set *Minpt* as 5 for all cases. The value of *Epsilon* is set differently for each airport pair to find the best result that matches the distribution pattern. We compute the $k^{th}$-nearest neighbor distances – where $m$ equals *Minpt* – plot these $k^{th}$- distances in descending order for all data points, and find the first "valley" of the sorted $k^{th}$-distance graph (Ester et al., 1996). This "valley" point corresponds to a threshold point where a sharp change in the gradient occurs along the $k^{th}$-distance curve, representing a change in density distribution of data points. The $k^{th}$-distance value of this "valley" point is used as the *Epsilon* value for DBSCAN. Take the airport pair, Shanghai Hongqiao Airport (SHA) and Guangzhou Baiyun Airport (CAN) for instance, Fig. 7(a)-(b) shows the sorted $5^{th}$-distance distribution and the histogram of the $5^{th}$-distance. The red point is in Fig. 7(a) corresponds to the first valley point where the gradient changes significantly. The *Epsilon* value is set to 1.3 (Euclidean distance between resampled trajectories in normalized unit) in DBSCAN for clustering trajectories of this airport pair. Two clusters are identified as illustrated in Fig. 7(c), representing two typical operational air routes connecting SHA and CAN. More details on how to identify operational air routes from trajectory data can be found in our previous study (Ren and Li, 2018).

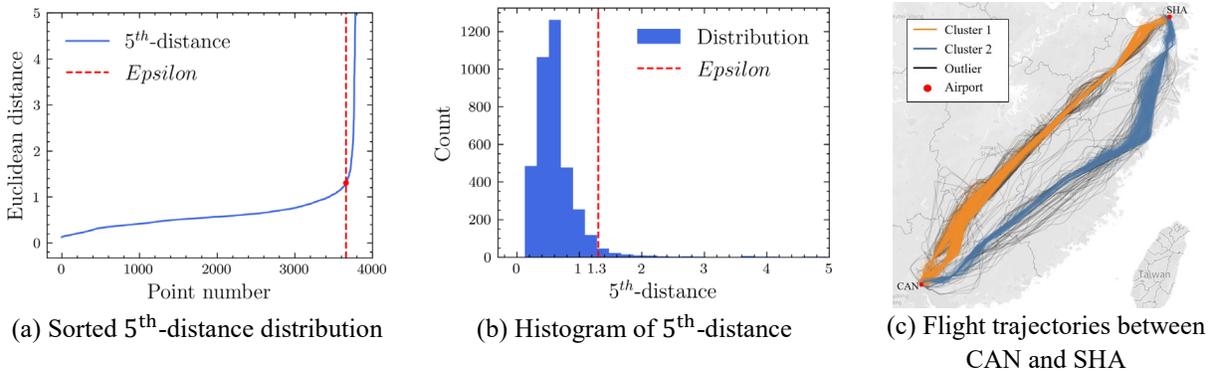

(a) Sorted $5^{th}$-distance distribution    (b) Histogram of $5^{th}$-distance    (c) Flight trajectories between CAN and SHA

Fig. 7 DBSCAN clustering for CAN-SHA airport pair.

In this study, by applying DBSCAN on the dataset described in Section 2.1, a total of 1376 air routes linking the 56 busiest airports in China are identified from 169140 trajectories, as shown in Fig. 8.

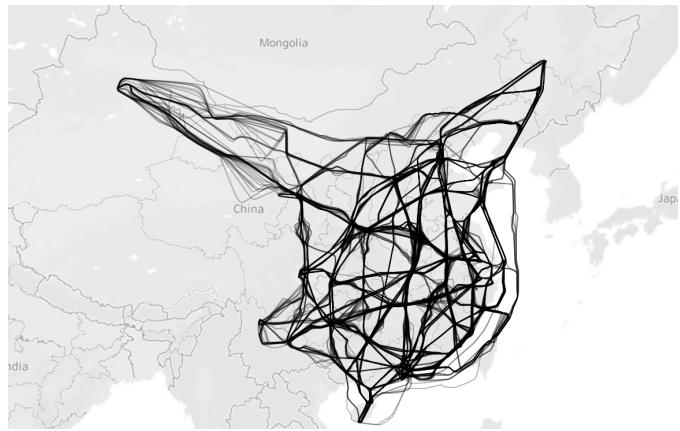

Fig. 8 Identified operational air routes in China.

### 2.2.2 En-route congestion point identification

In most existing studies, the en-route airspace is represented by a set of en-route sectors in an ATM system. However, sector-based en-route airspace models require extensive input information, i.e., sector maps. Furthermore, it is difficult to estimate the sector capacity accurately, because their characteristics and ATC tactics vary across different sectors and controllers. To tackle these challenges, our approach is to identify conceptual en-route congestion points from the aircraft tracking data directly. The conceptual en-route congestion points are defined as en-route airspace areas that have a high risk of air traffic congestion. Existing literature suggests how many air traffic controllers can handle in a sector is determined by air traffic complexity, which is measured by aircraft density or dynamic density (Hilburn, 2004; Mogford et al., 1995), the topological Kolmogorv entropy of the air traffic (Delahaye and Puechmorel, 2000; Prandini et al., 2011), etc.

We propose a data-driven algorithm to identify the conceptual en-route congestion points according to the principles discussed above. First, the whole airspace is discretized into cells by Cartesian grid as shown in Fig. 9. The operational flight routes recognized in the previous step are plotted as gray lines, while the cells used by these flight routes are represented as red rectangles. In this work, the grid size is defined as 20NM. The grid size can be adjusted within a range, where the upper limit is a typical size of an en-route sector (50 – 100 NM), and the lower limit is constrained by the computational time as a result of the total number of cells for the area of interest.

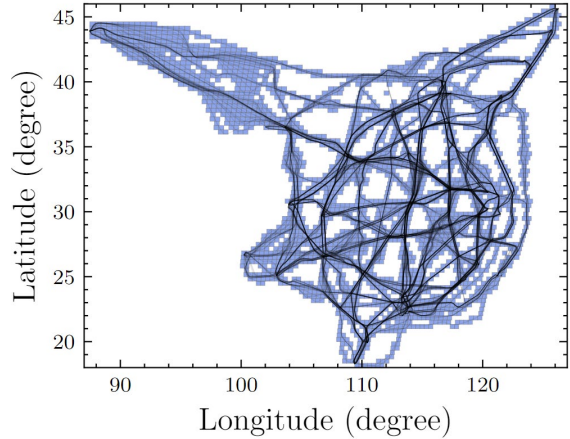

Fig. 9 Operational flight routes and mesh grids.

Then, three metrics are formulated to evaluate the risk level of congestion of each cell: the traffic load, the number of operational routes, and the entropy of route directions. Afterward, a score function is constructed by combining these three metrics. A cell with a higher score indicating a higher traffic load and a more complex route structure in this region, which is more likely to reach its capacity. Aircraft that fly through these regions have a high chance of experiencing delays or diversion. For the $i$th cell of the airspace, its score is computed as follows:

$$Score_i = \omega_1 \tilde{T}_i + \omega_2 \tilde{R}_i + \omega_3 \tilde{E}_i \tag{1}$$

$$E_i = -\sum_{j}^{N} p_j \log p_j \tag{2}$$

$$\tilde{T}_i = \frac{T_i - T_{min}}{T_{max} - T_{min}}, \quad \tilde{R}_i = \frac{R_i - R_{min}}{R_{max} - R_{min}}, \quad \tilde{E}_i = \frac{E_i - E_{min}}{E_{max} - E_{min}} \tag{3}$$

where $\{\tilde{T}_i, \tilde{R}_i, \tilde{E}_i\}$ are the standardized values of $\{T_i, R_i, E_i\}$, respectively, using Eq. (3) for standardization. $T_i$ represents the total traffic load in the $i$th cell, $R_i$ denotes the number of operational routes in the $i$th cell, and $E_i$ is the entropy in the $i$th cell. $\omega_1, \omega_2$, and $\omega_3$ are the corresponding coefficients. $(T_{min}, T_{max})$, $(R_{min}, R_{max})$, and $(E_{min}, E_{max})$ are the minimum and maximum values of all $T_i, R_i$, and $E_i$, respectively. $E_i$ evaluates the diverseness of route directions in this cell, and it is computed using Eq. (2), where $N$ represents the total number of flight directions within this cell and $p_j$ is the probability that aircraft flies in the $j$th direction, which can be inferred using flight trajectory data. The detailed calculation of $E_i$ is explained using Fig. 10.

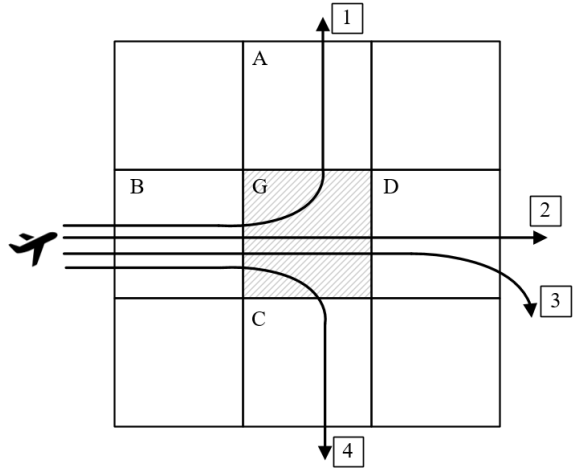

Fig. 10 Illustration of flight direction and grids.

In this example, the airspace is divided into 9 square cells as shown in Fig. 10. There are four flight routes passing cell $G$, namely, flight routes 1, 2, 3 and 4. Four cells share the same border with cell $G$, i.e., cell $A, B, C$, and $D$. We use cells that before entering and after exiting cell $G$ to identify the direction for these flight routes. For example, flight route 1 passing through cells $\{B, G, A\}$ sequentially. Thus, the first direction for cell $G$ is $d_{BA}$. Likewise, we can find three flight directions in cell $G$, that is, $\{d_{BA}, d_{BD}, d_{BC}\}$, denoted as direction $\{1, 2, 3\}$, respectively. So, $N = 3$ for this case. Moreover, let the traffic load for each flight route be 100. Therefore, the total traffic load $L_{sum}$ of cell $G$ is 400 and the traffic load for each direction $L_j = \{100, 200, 100\}$, $j \in \{1, 2, 3\}$. Finally, the probability of each flight direction $p_j$ and the entropy for this cell $E_G$ can be computed using Eq. (2). Specifically, $p_j = \{0.25, 0.5, 0.25\}$, and $E_G = 3.47$ for this example. The entropy value $E$ for all the cells in the whole space are calculated using the method discussed above, and then $\tilde{E}_G$ denoted as the standardized value of $E_G$ is computed according to Eq. (3).

We observe that the heat maps for the number of flight routes and traffic load follow similar patterns, while the heat map of entropy is different. Therefore, we fix $\omega_1 = \omega_2 = 1$, and iterate $\omega_3$ over $\{0.4, 0.6, 0.8, 1, 2, 3\}$ to compute the scores of the cells of the whole airspace. Results show that although the score values may be different with different combinations of $\{\omega_1, \omega_2, \omega_3\}$, those cells with a high score value are quite stable. We choose the top 75 cells that are insensitive to different settings of parameters as the cells at a high risk of congestion. The heat map of the whole airspace when $\{\omega_1 = 1, \omega_2 = 1, \omega_3 = 2\}$ and the standardized score value distribution is shown in Fig. 11. Fig. 11 (a) shows the spatial distribution of the cells with different score values; a higher score is shown with denser color. Fig. 11 (b) depicts the score value distribution and the turning point is around 0.6. Cells with score values larger than 0.6 are considered as with a high level of congestion risks. These cells are further clustered into 30 clusters as conceptual en-route congestion points, as shown in Fig. 12 (a). In this step, DBSCAN is used to perform the clustering, and the *Minpt* and *Epsilon* are set as 2 and 50NM, respectively. A sensitivity analysis was performed to select the DBSCAN parameters. We found that results are not sensitive to different settings of parameters. The number of clusters found is stabilized around 30.

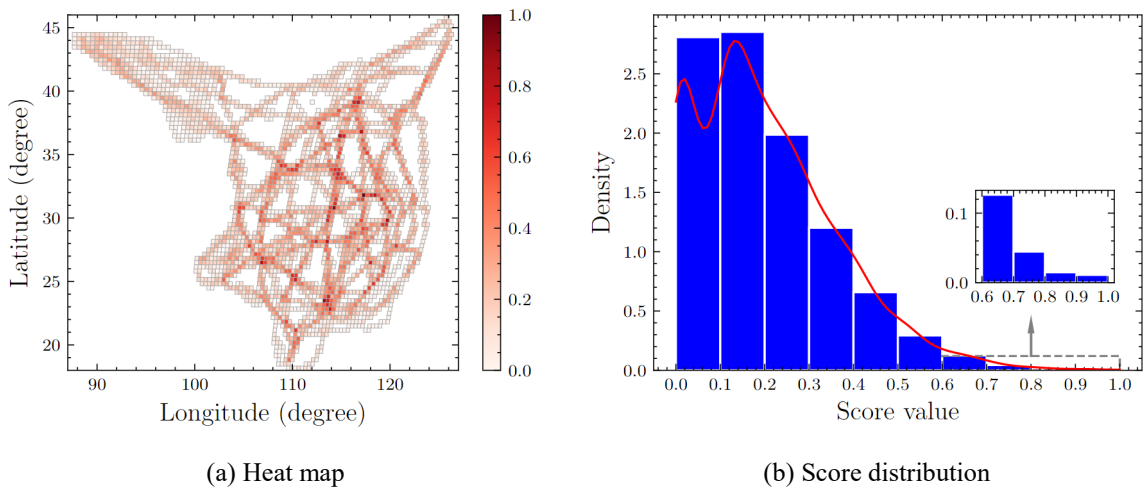

(a) Heat map    (b) Score distribution

Fig. 11 Heat map and score distribution of mesh grids ($\omega_1 = 1, \omega_2 = 1, \omega_3 = 2$).

The 30 conceptual en-route congestion points identified via this approach were compared with the busiest en-route waypoints published in CAAC (2016), as shown in Fig. 12. The identified en-route congestion points are depicted by yellow circles, while the top 10 busiest waypoints published by CAAC are indicated by red triangles. More information of the top 10 busies waypoints are shown in Table 1. Seven out of the top ten busiest waypoints were recognized as the en-route congestion points in our method. These waypoints have an average of 900 - 1800 flights passing through each day. Among these waypoints, WXI, ZHO and LKO are located on high-density trunk route, A461 linking Beijing and Guangzhou. HFE and MADUK are where the trunk route R343 and other routes converge serving many airports in Jiangsu and Shanghai area. LLC and MAMSI are where several high-density air routes converge. Three out of the top ten busiest waypoints were not recognized by our method, i.e., OBLIK, PAVTU, and PLT. The potential reason could be that the traffic flows are not complicated at these waypoints. There is one major route passing through OBLIK, two major routes at PAVTU, and one major route at PLT.

The en-route congestion points are not the same as the busy waypoints, as the former are evaluated using three metrics discussed above, while the latter are based on traffic count only. Many studies have pointed out simple air traffic count is not a good measure for airspace capacity - a sector can have a high level of traffic but not congested if the traffic flows are organized and independent. Airspace capacity is determined by air traffic controller workload and air traffic complexity, which are measured by aircraft density, dynamic density, the level of disorder and the organization structure of the air traffic distribution measured by aircraft velocities and positions, the topological Kolmogorv entropy of air traffic, etc. (Delahaye and Puechmorel, 2000; Hilburn, 2004; Mogford et al., 1995; Prandini et al., 2011). We used a similar metric to determine possible congestion points, including traffic count, flight route number, and entropy. Note that en-route airspace with temporary congestion, such as weather impact, temporary airspace restrictions, are not recognized in this method.

Table 1 Top 10 busiest waypoints published by CAAC (CAAC, 2016).

| Waypoint name | Daily traffic | Recognized by our method |
|---|---|---|
| Zhoukou (ZHO) | 1726 | Yes |
| Luogang (HFE) | 1649 | Yes |
| Weixian (WXI) | 1480 | Yes |
| OBLIK | 1185 | No |
| MADUK | 1138 | Yes |
| Longkou (LKO) | 1077 | Yes |
| PAVTU | 1033 | No |
| MASMI | 933 | Yes |
| Laoliangcang (LLC) | 922 | Yes |
| Panlong (PLT) | 860 | No |

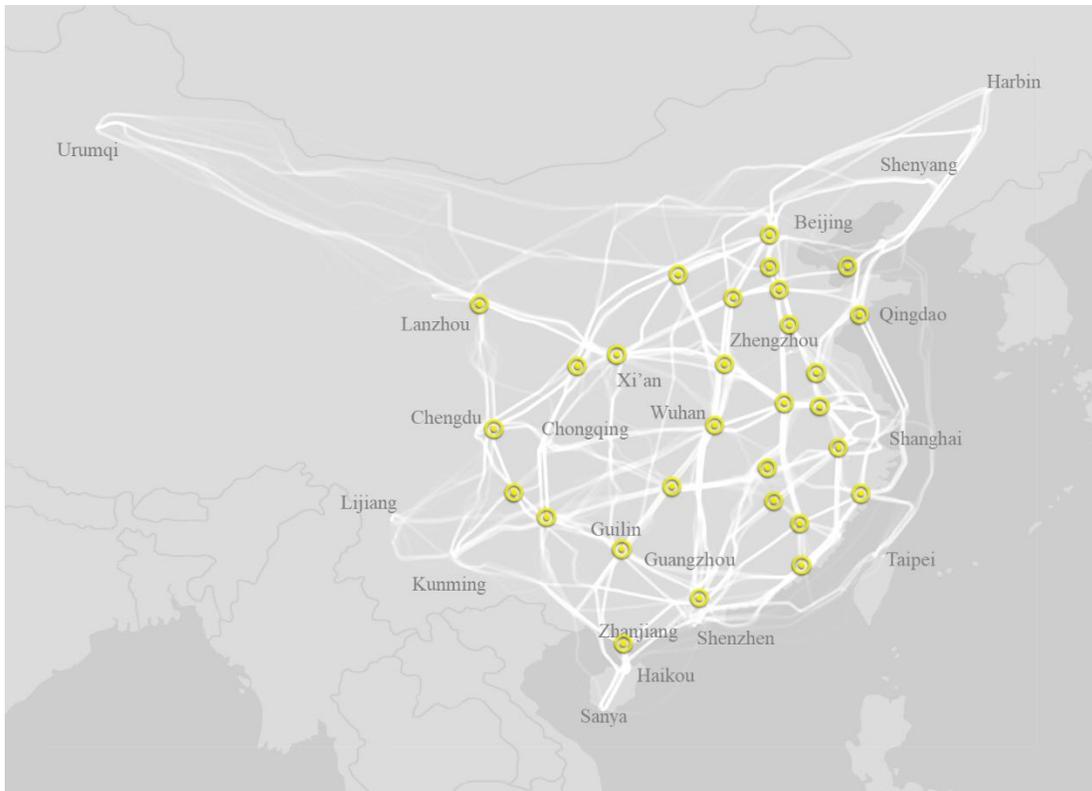

(a) En-route congestion points identified via the proposed method

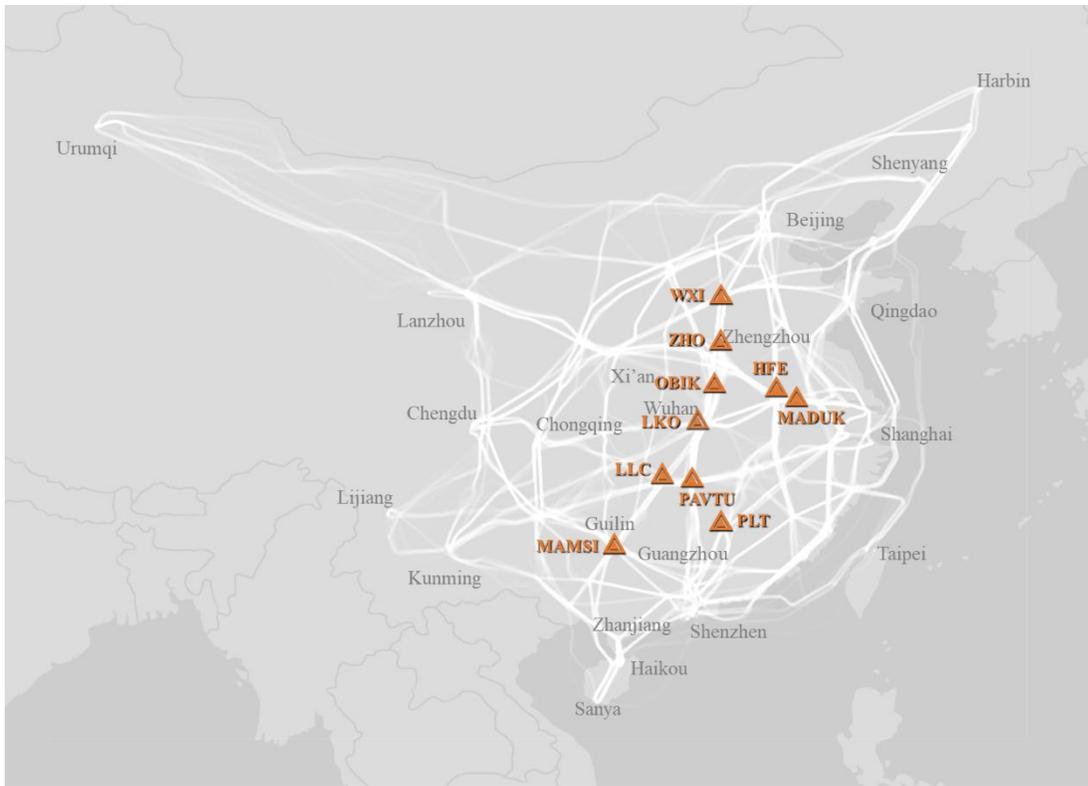

(b) Top 10 busiest waypoints published by CAAC (CAAC, 2016)

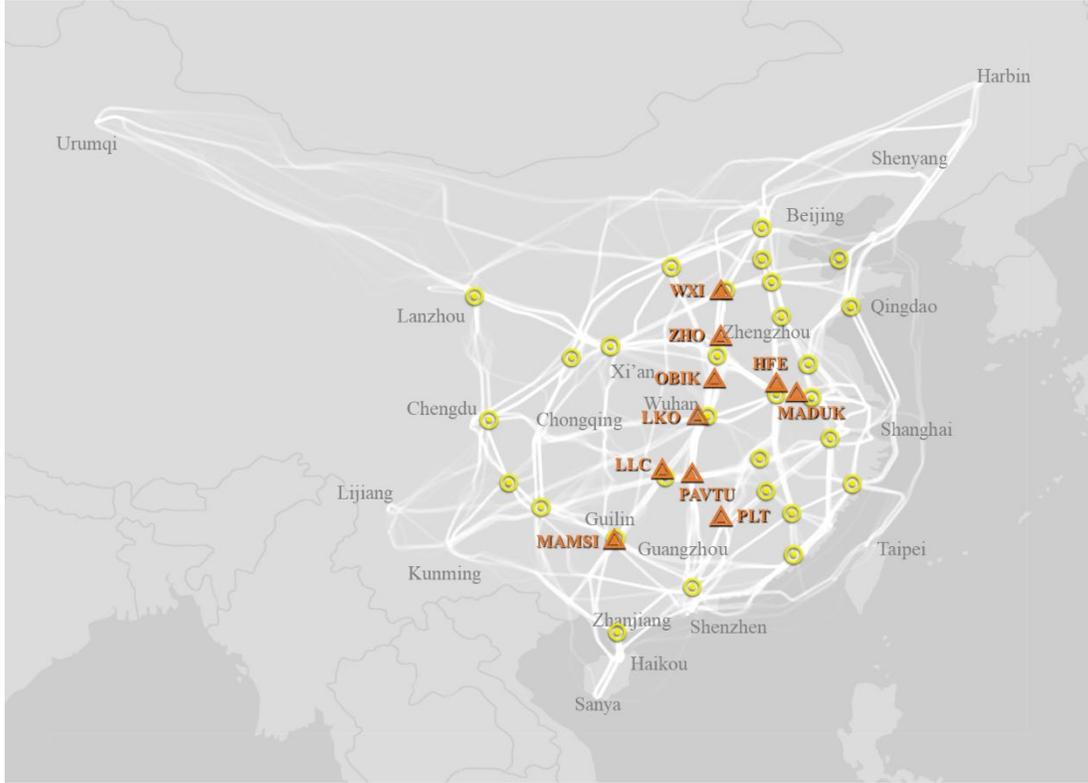

(c) Comparison of en-route congestion points and busy waypoints

Fig. 12 Identified en-route congestion points and top 10 busiest waypoints.

## 2.3 A stochastic and dynamic queuing network model to calculate delays and track their propagation

The second part of the proposed framework involves a stochastic and dynamic queuing network model that treats each airport or each en-route congestion point as a node of the queuing system to model flight delays and their propagation. The core concept of the proposed Multi-layer Air Traffic Network Delay (MATND) model is similar to the AND model proposed by Pyrgiotis et al. (2013), which combines a numerical queuing engine (QE) to compute the approximate delays at each airport or en-route congestion point with a delay propagation algorithm (DPA) to track the propagation of these delays from one node to another over the period of interest.

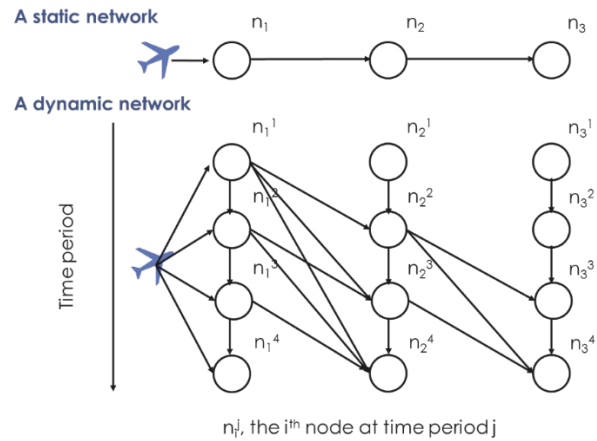

Fig. 13 Time-space network concept in MATND.
(A node is an airport or a conceptual en-route congestion point.)

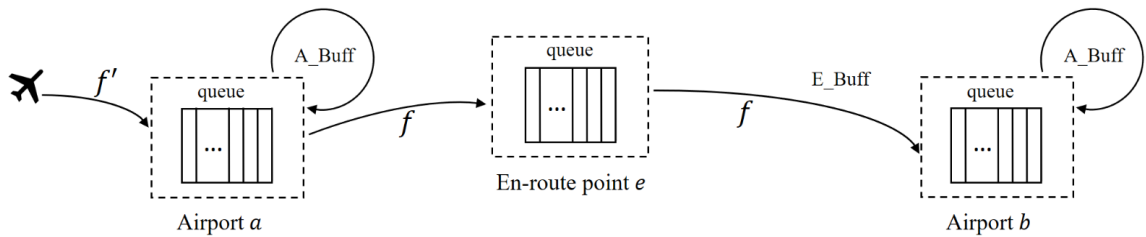

Fig. 14 A simplified schedule for an aircraft.

MATND considers the air traffic system as a dynamic network. Flight delays on this network are calculated using the time-space decomposition method. The concept of the time-space decomposition method is shown in Fig. 13, where we can see the dynamic network is different from the static one as the nodes in the network are connected by flights at different periods, and the characteristics (e.g., service profiles, demand profiles) at each node can be adjusted during the iteration of the network. For better illustration, a simplified schedule for an aircraft is given as Fig. 14. There are two flights $\{f', f\}$ in this example, where $f'$ is the predecessor flight of $f$. The aircraft first arrives at airport $a$ and finishes its flight $f'$. Then the aircraft starts its next flight $f$, departing from airport $a$, then passing through the en-route point $e$ and finally arriving at airport $b$. MATND models the airport as a single server that serves both arrivals and departures, and the en-route point as a single server that serves passing through airplanes.

The flow chart of MATND is shown in Fig. 15. MATND computes flight delays over a network for a period $T$ (usually 24 h) via iterative calculations on sub-periods, $h_j$, $j = 1, 2, ..., m$, i.e., $m = 96$, each of which is 15 min long ($\Delta T$). Starting at the beginning time $h = T_0$, the flights $F_{a,h}$ and $F_{e,h}$ at airport $a$ and en-route point $e$, respectively, are collected. QE is then executed for each flight $f \in \{F_{a,h}, F_{e,h}\}$ individually, and the expected delay $W_a(t)$ and $W_e(t)$ of each flight $f$ are estimated. The DPA then determines whether the delay incurred by any flight is significant enough to result in propagation. If so, the earliest time $t^*$ at which this propagation condition is met is identified. All flight operations terminated before $t^*$ are unaffected; these flights are classified as processed and are not affected by any subsequent iterations of the algorithm. For all unprocessed flights affected by the propagation, DPA adjusts their arrival and departure times, and the demand rates at corresponding airports are updated accordingly. The QE is rerun at each node based on the updated demand rates. The QE-DPA process is executed iteratively until $h$ reaches the end of time $T_{end}$.

It is worth noting that QE runs on a finer-grain time step than $\Delta h$. Therefore, we denote the instantaneous demand and service rates used by the QE as $\lambda(t)$ and $\mu(t)$, respectively, where $t \in \mathfrak{R}, T_0 \leq t \leq T_{end}$. Also, a $index$ function is formulated to map time $t$ into the sub-period $h_j$. The $index(t)$ is given by $j = index(t) = \left\lfloor \frac{t}{\Delta T} \right\rfloor$.

The assumptions used in the MATND model are as follows: (1) No flight can depart earlier than its scheduled departure time; (2) Flights can arrive at the destination airport before their arrival time; (3) En-route delays occurring at a congestion point cannot be pushed back to upstream en-route points or the origin airport; (4) Flight cancellations are not incorporated into the model; and (5) Ground delays are not considered in the model.

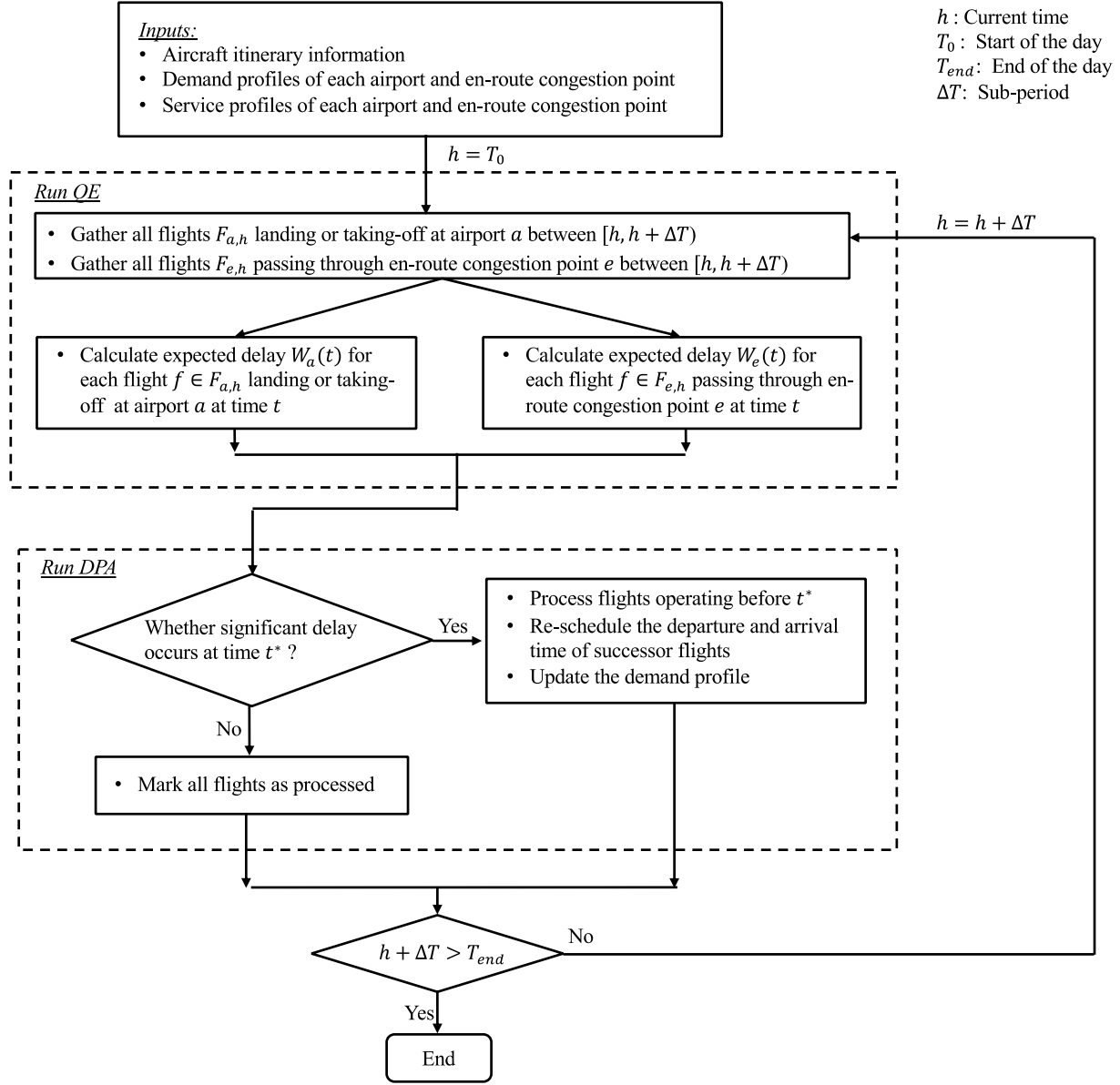

Fig. 15 Delay calculation flow chart of MATND.

### 2.3.1 Modeling each airport as a queuing system

The aircraft arrival and departure processes at the runway system of an airport are simulated using queuing models with a nonstationary Poisson arrival process, time-dependent $k_r$ th-order Erlang service-time distribution, a single server, the first-come-first-served (FCFS) service principle, and an infinite waiting room (we denote this model type as $M(t)/E_{k_r}(t)/1$) (Malone, 1995; Pyrgiotis et al., 2013).

The expected waiting time at a particular time of such a queuing model is calculated using a numerical queuing engine, DELAYS, developed by Kivestu (1976) and Malone (1995). The state-transition diagram of this queuing system consists of $(k_r N + 1)$ stages, where $N$ is the queue capacity (Larsen and Odoni, 1981). To approximate such an infinite-capacity system, $N$ must be sufficiently large that the probability of having $N$ or more customers in the system is very small. As the number of equations $(k_r N + 1)$ to be solved in the queuing system becomes very large, finding the exact solution requires significant amounts of computer memory and CPU time. For this reason, we use an approximation of the $M(t)/E_{k_r}(t)/1/N$ system that solves a set of $(N + 1)$ difference equations, instead of the $(k_r N + 1)$ Chapman–Kolmogorov equations (Kivestu, 1976). Extensive computational experiments performed by Malone (1995) indicate that Kivestu's approach approximates the numerical solution of $M(t)/E_{k_r}(t)/1$ very accurately and is much faster than the exact method.

The expected waiting time, $W_a(t)$ at time $t$ of airport $a$, is calculated as follows:

$$W_a(t) \approx \frac{L_a(t)}{\mu_a(t)}, \tag{4}$$

$$L_a(t) = \sum_{j=1}^{N} (j - 1) p_{a,j}(t), \tag{5}$$

where $L_a(t)$ is the expected number of aircraft in the queue of airport $a$ at time $t$, $\mu_a(t)$ is the service rate of airport $a$ at time $t$, and $N$ is the capacity of the queue. $p_{a,j}(t)$ denotes the estimated state probability, more specifically, the probability of having $j$ airplanes in airport $a$'s queuing system at time $t$. The state probabilities are estimated using Kivestu's approximation approach (Kivestu, 1976).

$$p_j(t_{l+1}) = p_0(t_l) \alpha_{l+1}(j) + \sum_{i=1}^{j+1} (j - 1) p_i(t_l) \alpha_{l+1}(j - i + 1), \tag{6}$$

$$j = 0, 1, \ldots, N$$

$$t_{l+1} = t_l + \left(\frac{k_r + 1}{k_r}\right)\left(\frac{1}{\mu(t_l)}\right), \tag{7}$$

$$\alpha_{l+1}(r) = \frac{\left(\frac{\lambda(t_l)}{\mu(t_l)}\right)^r e^{-\frac{\lambda(t_l)}{\mu(t_l)}}}{r!}, \tag{8}$$

where $t_l$ is the epoch time of $l^{th}$ aircraft completes the arrival/departure process. Given initial conditions for the system at time $t_0$, we can compute the state probabilities of any epoch $t_l$.

The initial state of the queuing system affects the expected waiting time. Normally the system is assumed to have an empty state at the beginning of the day. According to the aircraft tracking data we collected, the whole airspace is almost empty at 4:00 AM (HKT). Therefore, in this study, the modeling horizon starts at $T_0 = 4{:}00$ AM and ends at $T_{end} = 4{:}00$ AM on the following day. The mathematical formulation of an empty state is given as Eq. (9).

$$p_j(0) = \begin{cases} 1, & j = 0 \\ 0, & j = 1, 2, \ldots, N \end{cases} \tag{9}$$

We use the method adopted by Malone, K.M. (1995) to derive $k_r$ for each airport. Firstly, the service time of each airport is collected, and then the mean, $\bar{x}_e$, and variance $\sigma_e^2$ are computed for the service time. Then we can find $k_r$ for an Erlang distribution to approximate the actual service time distribution as follows:

$$\frac{(\bar{x}_e)^2}{k_r} = \sigma_e^2 \Rightarrow k_r = int\left(\frac{(\bar{x}_e)^2}{\sigma_e^2}\right), \tag{10}$$

where $int()$ represents the function that converts the result to an integer.

The initial values of $\lambda(t_l)$ are set based on the demand profile of the target interval at each airport, which is estimated through the scheduled aircraft itinerary information. The service rate $\mu(t_l)$ is simplified to be a fixed number, the capacity of an airport, in this study. Fig. 16 shows the demand and service profiles at Beijing Capital International Airport (PEK) over a single day. The demand profile reflects the number of scheduled flights per 15 min, whereas the service profile reflects the number of served flights per 15 min. The initial value of $\lambda(t_l)$ of each airport is defined as the expected number of scheduled arrivals and departures in each sub-period. The service rate $\mu(t_l)$ of each airport is estimated empirically based on historical data. We count the number of flights served (including arrivals and departures) per 15 min for each airport within a one-month period, which includes 2880 sub-periods. The service rate is conservatively defined as the lower bound of the number of flights served that covers 90% of the sub-periods in the monitored period. Fig. 17 shows the demand and service rates at PEK airport based on operational flight data for one month.

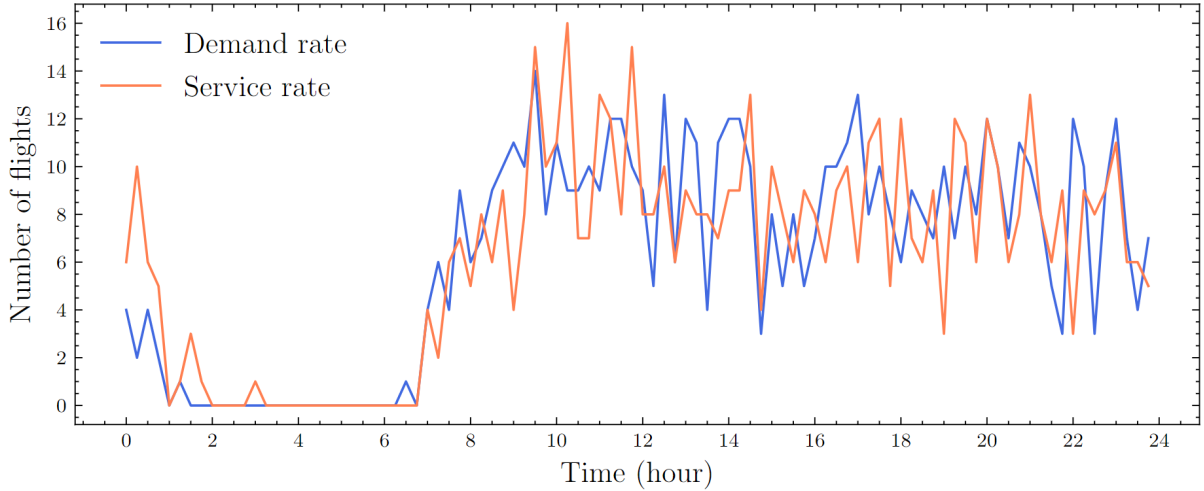

Fig. 16 Demand profile and service profile at PEK during a single day.

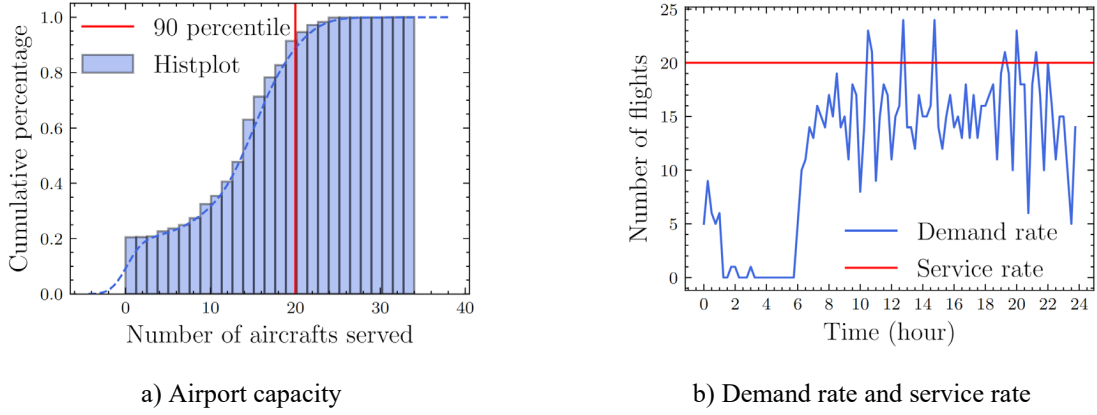

a) Airport capacity          b) Demand rate and service rate

Fig. 17 Demand rate and service rate at PEK based on flight scheduling data for one month.

### 2.3.2 Modeling each en-route congestion point as a queuing system

In the proposed model, we treat the process of an aircraft passing through an en-route congestion point as a queuing process. This is because when an en-route sector is congested, flights may be instructed by air traffic controllers to slow down, enter a holding pattern, stay in upstream sectors (even receive a ground delay at their origin airport), or divert to a different route. Except for the route diversion, all other en-route congestion effects can be modeled as a queueing system, similar to the congestion effects at airports.

Each en-route congestion point is modeled as an $M(t)/E_{k_r}(t)/1$ queuing model, which is similar to the queuing system applied at the airports. The expected waiting time of en-route congestion point $e$ at time $t$, $W_e(t)$, and the expected number of aircraft in the queue of en-route congestion point $e$ at time $t$, $L_e(t)$, are calculated as follows:

$$W_e(t) \approx \frac{L_e(t)}{\mu_e(t)}, \tag{11}$$

$$L_e(t) = \sum_{j=1}^{N}(j-1)p_{e,j}(t), \tag{12}$$

where $p_{e,j}(t)$ is the probability of having $j$ airplanes at en-route congestion point $e$ of the queuing system at time $t$, and $\mu_e(t)$ is the service rate at time $t$. The state probabilities are estimated in the same way as the airport queuing system, using Kivestu's approximation approach (Kivestu, 1976), as described in Section 2.3.1.

The demand profile at each en-route congestion point cannot be generated directly from the flight schedule data, which only contains the flight information at airports. Using the multi-layer air traffic network constructed in Section 2.2.2, given the demand profile of each airport and the usage probability of each air route, we can obtain the demand profile of each air route. The average time required from each origin airport to the en-route point along a specified air route is estimated based on the historical aircraft tracking data. Together with the demand profile of each air route, we estimate the demand profile at each en-route congestion point. The service profile is obtained from the ADS-B aircraft trajectory data directly. Fig. 18 shows the demand and service profiles at an en-route congestion point during a single day. The demand and service rates are estimated by the same method used for the airport queuing system. Fig. 19 indicates the demand rate and service rate at a randomly selected en-route congestion point.

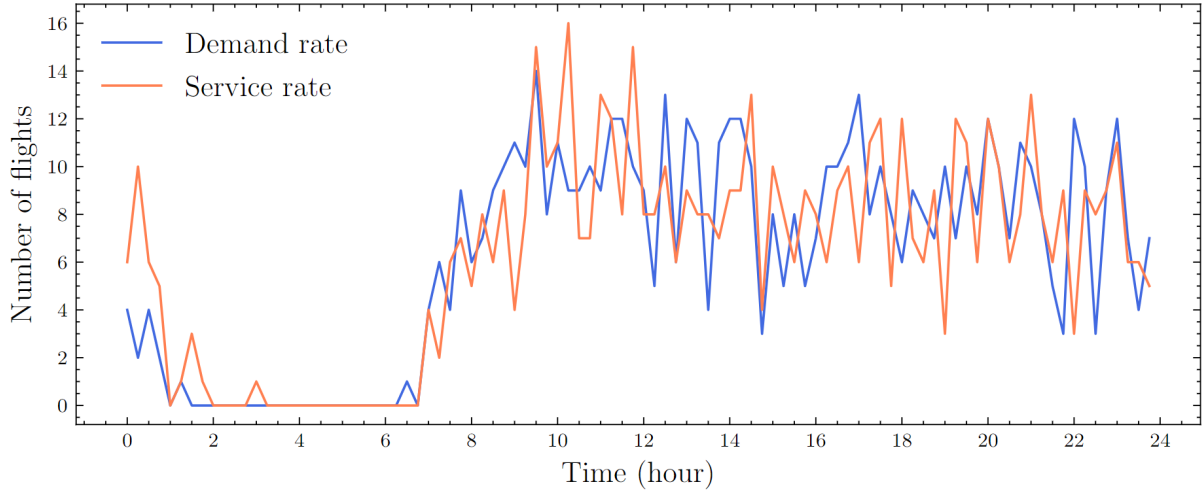

Fig. 18 Demand profile and service profile at an en-route congestion point during a single day.

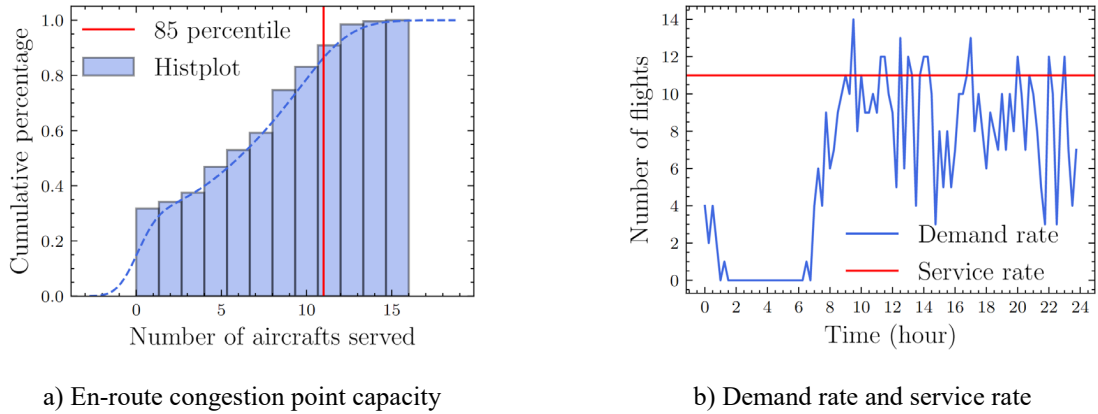

a) En-route congestion point capacity

b) Demand rate and service rate

Fig. 19 Demand rate and service rate at an en-route congestion point based on flight scheduling data.

### 2.3.3 Delay propagation through the air traffic network

The Delay Propagation Algorithm (DPA) propagates significant delays generated at any node of the network (both airports and en-route congestion points) to the "downstream" itinerary of this aircraft. We adopted the same DPA as in the AND model (Pyrgiotis et al., 2013) except that we added en-route congestion points as nodes that may generate local delays and propagate significant delays in the network. The difference between DPA in AND and the one in MATND is illustrated in Fig. 20. DPA in MATND incorporates the en-route points into the algorithm. Thus, delays can be generated at these en-route points and propagated to downstream flights as well.

The main idea of DPA can be described as follows:

1) Determine if the delay is significant enough to propagate downstream.

2) Propagates significant delays to the downstream itinerary of an aircraft (the itinerary of an aircraft consists of both airports and en-route congestion points).

3) Re-schedule the arrival and departure times of delayed flights.

4) Update the demand rates of all airports and en-route congestion points.

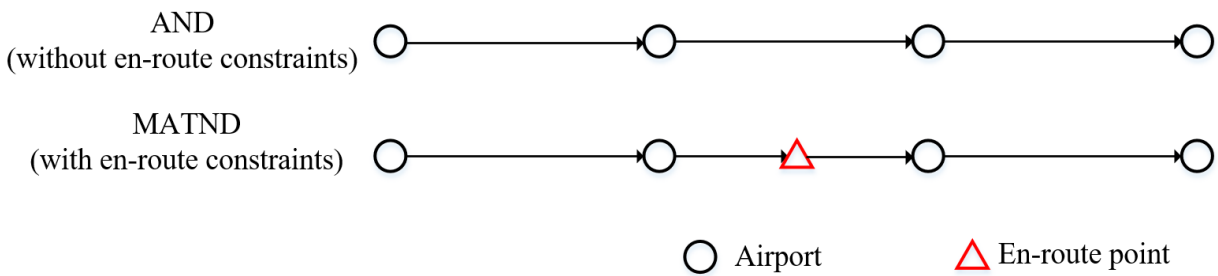

Fig. 20 Conceptual illustration of DPA in AND and DPA in MATND.

Whether a delay is significant or not is determined by two important parameters in DPA, ground time buffer (A_Buffer) and en-route buffer (E_Buffer) shown as Fig. 14. We adopted the same definition of ground time buffer as it is used in the AND model, while added the en-route buffer in our method. If we denote the immediate predecessor flight of each flight $f$ as $f'$, the parameter A_Buffer indicates the ground time buffer between two consecutive flights of the same aircraft at an airport, which could absorb flight delays at the airport. Flight $f$ will still depart on time as long as the arrival of $f'$ is delayed by less than A_Buffer. The en-route buffer E_Buffer is the buffer between the scheduled flight time and the actual flight time. In actual operations, even though a flight may suffer a departure delay, it can still arrive on time or even early because the scheduled flight time issued by airlines is longer than the actual flight time. The en-route buffer can, therefore, absorb departure delays. Delays that cannot be absorbed by A_Buffer and E_Buffer are recognized as significant delays, and the downstream itinerary executed by the aircraft must be adjusted. In this study, A_Buffer and E_Buffer are set to fixed values for all flights. More details on DPA calculations can be found in the paper by Pyrgiotis et al. (2013).

## 3 Theoretical comparison between MATND and AND on a stylized network

In this section, we theoretically analyze the advantages of the proposed MATND model in comparison with the classic network delay model, AND, under stylized assumptions. Since it is ex-tremely complicated to conduct a theoretical analysis over a general network, we construct a simple network with three airports and an en-route congestion point to conduct such analysis, as shown in Fig. 21 and Fig. 22. In this network, there are three airports, i.e., A1, A2 and A3. A1 is the source node, generating a number of flights flying to A2. Some of these flights and together with a number of new flights then depart from A2 to their final destination A3. There might be congestions at the runway system of airport A2 for both arrival and departure, as well as congestions in the en-route airspace between A1 and A2. The focus of this theoretical analysis are the arrival delay of flights and the departure delay of flights at A2. We show that AND cannot be turned to approximate both the arrival delay and the departure delay at the same time while MATND can due to their network structure differences.

In the AND model, the arrival delay of flights and the departure delay of flights at A2 are considered to be generated at the runway queuing system of A2, which is an $M/G/1$ system. Flights from A1 to A2 ($f'$) generate demands at A2 for arrival, which follow a Poisson process with an average rate of $\lambda_1$. Flights from A2 to A3 ($f$) generate departure demands at A2, which also follow a Poisson process with an average rate of $\lambda_2$. Both arrivals and departures at A2 are served by one runway system with equal priority according to a FCFS discipline.

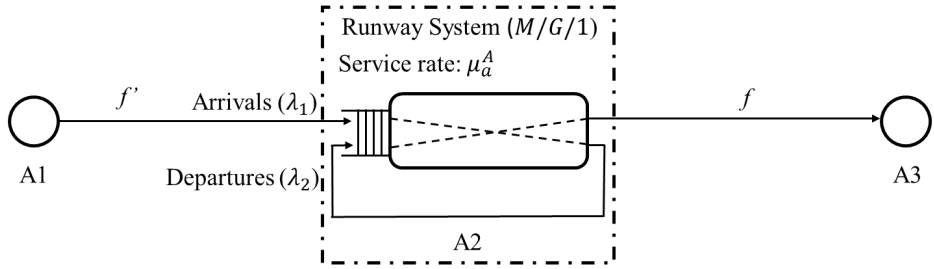

Fig. 21 Network structure of the stylized network of the AND model, where the runway system of airport A2 is illustrated.

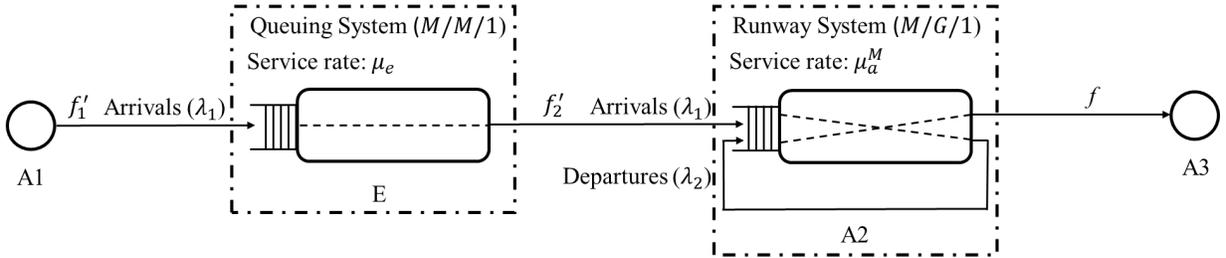

Fig. 22 Network structure of the stylized network of the MATND model, where the queuing system of en-route point and the runway system of airport A2 are illustrated.

In the MATND model, in addition to the runway queuing system of A2, an en-route congestion point E is inserted between A1 and A2 to capture the en-route delays. In this stylized example, we model the en-route congestion point E as an $M/M/1$ queuing system, which is a special case of $M/G/1$, to make the theoretical proof easier. Flights $f'$ are divided into two segments, $f_1'$ and $f_2'$, as the input and output of the queuing system at en-route congestion point E. Due to the special property of $M/M/1$, both $f_1'$ and $f_2'$ follow a Poisson process with an average rate of $\lambda_1$. The rest of the queuing network structure in the MATND model remains the same as in the AND model.

To summarize, the only difference between the AND and MATND is whether the en-route delay is captured by another queuing node (in addition to the airport queuing nodes) or not. The network structures for the AND and MATND models are shown in Fig. 21 and Fig. 22, respectively.

The dynamic behavior of queuing system is complex. There is no closed-form solution for the time-dependent system. In the stylized example, we compare the performance of the AND and MATND by deriving the key statistical measures of the system's long-term equilibrium behavior. Specifically, in the AND model, when the system reaches its steady state, the expected value of the arrival delay of flights from A1 to A2 in the AND model is computed by:

$$E[D_{a,\text{arr}}^A] = E[AA(f') - SA(f')] = E[W_a^A]$$
$$= E[W_{a,q}^A] + \frac{1}{\mu_a^A} \quad (13)$$

where the subscript $a$ represents airport A2, the subscript arr represents the arrival process of airport A2, and the superscript $A$ represents the AND model. The actual arrival time and the scheduled arrival time of flight $f'$ are denoted by $AA(f')$ and $SA(f')$, respectively. The total amount of time spent by flight $f'$ in the queuing system of A2 is denoted by $W_a^A$, which consists of two components, i.e., the queue waiting time $W_{a,q}^A$, and the service time with an expected value $\frac{1}{\mu_a^A}$.

Based on the Pollaczek-Khinchine (P-K) formula (Khintchine, 1932; Pollaczek, 1930), the expected waiting time in queue for the $M/G/1$ system is:

$$E[W_{a,q}^A] = \frac{(\rho_a^A)^2 + (\lambda_1 + \lambda_2)^2 \sigma^2(S_a^A)}{2(\lambda_1 + \lambda_2)(1 - \rho_a^A)} \quad (14)$$

where $\lambda_1$ is the arrival rate of airport $a$, $\lambda_2$ is the departure rate of airport $a$, $\mu_a^A$ stands for the service rate of airport $a$, $\rho_a^A$ is the utilization, and $\sigma^2(S_a^A)$ is the variance of the service time distribution. In both the AND and the MATND, a Erlang distribution with order $k_r$ is used to approximate the service process at the airport. Therefore, we have:

$$\rho_a^A = \frac{\lambda_1 + \lambda_2}{\mu_a^A} \quad (15)$$

$$\sigma^2(S_a^A) = \frac{1}{k_r(\mu_a^A)^2} \quad (16)$$

Likewise, the expected departure delay of flights from A2 to A3 for the AND model, denoted by $E[D_{a,\text{dep}}^A]$, is calculated by:

$$E[D^A_{a,\text{dep}}] = E[AD(f) - SD(f)] = E[W^A_a]$$
$$= E[W^A_{a,q}] + \frac{1}{\mu^A_a} \tag{17}$$

where $AD(f)$ and $SD(f)$ are the actual departure time and scheduled flight time of flight $f$, respectively.

In the MATND model, the arrival delay of flights from A1 to A2 is the sum of two parts: the airport delay $W^M_a$ and the en-route delay $W_e$. Therefore, we have:

$$E[D^M_{a,\text{arr}}] = E\left[AA(f') - SA(f')\right] = E[W^M_a] + E[W_e]$$
$$= E[W^M_{a,q}] + \frac{1}{\mu^M_a} + E[W_{e,q}] + \frac{1}{\mu_e} \tag{18}$$

The service process at the airport A2 is approximated as a Erlang distribution with order $k_r$ and average service rate $\mu^M_a$. At the en-route congestion point E, the service process is modelled as a Poisson process with average service rate $\mu_e$. Following the P-K formula, we have the below equations for $E[W^M_a]$ and $E[W_e]$.

$$E[W^M_{a,q}] = \frac{(\rho^M_a)^2 + (\lambda_1 + \lambda_2)^2 \sigma^2(S^M_a)}{2(\lambda_1 + \lambda_2)(1 - \rho^M_a)} \tag{19}$$

$$\rho^M_a = \frac{\lambda_1 + \lambda_2}{\mu^M_a} \tag{20}$$

$$\sigma^2(S^M_a) = \frac{1}{k_r(\mu^M_a)^2} \tag{21}$$

$$E[W_{e,q}] = \frac{(\rho_e)^2 + \lambda_1^2 \sigma^2(S_e)}{2\lambda_1(1 - \rho_e)}. \tag{22}$$

$$\rho_e = \frac{\lambda_1}{\mu_e} \tag{23}$$

$$\sigma^2(S_e) = \left(\frac{1}{\mu_e}\right)^2 \tag{24}$$

The expected departure delay of flights from A2 to A3 for the MATND model is computed by

$$E[D^M_{a,\text{dep}}] = E[AD(f) - SD(f)] = E[W^M_a]$$
$$= E[W^M_{a,q}] + \frac{1}{\mu^M_a} \tag{25}$$

Let $D_{a,\text{arr}}$ denote the actual delayed time of arrival flights $f'$ at A2, and $D_{a,\text{dep}}$ denote the actual delayed time of departure flights $f$ at A2. In this stylized example, we assume not considering other exogenous delay factors, such as Ground Delay Program (GDP), airline reasons, etc. Therefore, $E[D_{a,\text{arr}}] \geq E[D_{a,\text{dep}}]$.

**Proposition 1.** MATND is able to accurately estimate the expected values of $D_{a,\text{arr}}$ and $D_{a,\text{dep}}$ simultaneously.

*Proof.* For MATND, based on Eqs. (19), (20), (21) and (25), we have

$$\begin{aligned}E[D^M_{a,\text{dep}}] &= f(\mu^M_a, k_r, \lambda_1, \lambda_2) \\ &= \frac{(\rho^M_a)^2 + (\lambda_1 + \lambda_2)^2 \sigma^2(S^M_a)}{2(\lambda_1 + \lambda_2)(1 - \rho^M_a)} + \frac{1}{\mu^M_a} \\ &= \frac{(k_r + 1)(\lambda_1 + \lambda_2)}{2k_r \mu^M_a (\mu^M_a - (\lambda_1 + \lambda_2))} + \frac{1}{\mu^M_a}\end{aligned} \quad (26)$$

We can tune $\mu^M_a$ to make $E[D^M_{a,\text{dep}}] = E[D_{a,\text{dep}}]$.

Then based on Eqs. (18)-(24), we have

$$\begin{aligned}E[D^M_{a,\text{arr}}] &= f(\mu^M_a, \mu_e, k_r, \lambda_1, \lambda_2) \\ &= \frac{(\rho^M_a)^2 + (\lambda_1 + \lambda_2)^2 \sigma^2(S^M_a)}{2(\lambda_1 + \lambda_2)(1 - \rho^M_a)} + \frac{1}{\mu^M_a} + \frac{(\rho_e)^2 + \lambda_1^2 \sigma^2(S_e)}{2\lambda_1(1 - \rho_e)} + \frac{1}{\mu_e} \\ &= \frac{(k_r + 1)(\lambda_1 + \lambda_2)}{2k_r \mu^M_a (\mu^M_a - (\lambda_1 + \lambda_2))} + \frac{1}{\mu^M_a} + \frac{1}{(\mu_e - \lambda_1)}\end{aligned} \quad (27)$$

With a given $\mu^M_a$, we can adjust $\mu_e$ to make $E[D^M_{a,\text{arr}}] = E[D_{a,\text{arr}}]$

Therefore, the proposed MATND is able to accurately estimate the expected values of $D_{a,\text{dep}}$ and $D_{a,\text{arr}}$ simultaneously.

**Proposition 2.** AND is not able to accurately estimate the expected values of $D_{a,\text{arr}}$ and $D_{a,\text{dep}}$ simultaneously.

*Proof.* For AND, based on Eqs. (13) - (16), we have

$$\begin{aligned}E[D^A_{a,\text{arr}}] &= f(\mu^A_a, k_r, \lambda_1, \lambda_2) \\ &= \frac{(k_r + 1)(\lambda_1 + \lambda_2)}{2k_r \mu^A_a (\mu^A_a - (\lambda_1 + \lambda_2))} + \frac{1}{\mu^A_a}\end{aligned} \quad (28)$$

Meanwhile, based on Eqs. (14) - (17), we also have

$$\begin{aligned}E[D^A_{a,\text{dep}}] &= f(\mu^A_a, k, \lambda_1, \lambda_2) \\ &= \frac{(k_r + 1)(\lambda_1 + \lambda_2)}{2k_r \mu^A_a (\mu^A_a - (\lambda_1 + \lambda_2))} + \frac{1}{\mu^A_a} \,.\end{aligned} \quad (29)$$

Function (28) and Function (29) have exactly the same formation, we have

$$E[D^A_{a,\text{dep}}] = E[D^A_{a,\text{arr}}]. \quad (30)$$

Therefore, we cannot tune $\mu_a^A$ to make $E[D_{a,\text{arr}}^A] = E[D_{a,\text{arr}}]$ and $E[D_{a,\text{dep}}^A] = E[D_{a,\text{dep}}]$ simultaneously unless $E[D_{a,\text{arr}}] = E[D_{a,\text{dep}}]$.

To further illustrate the impact on possible estimation bias of the AND, we consider the following two cases.

**Case 1.** Suppose we tune both models to make their estimated arrival delay of $f'$ equals to the actual value, $E[D_{a,\text{arr}}]$:

$$E[D_{a,\text{arr}}] = E[D_{a,\text{arr}}^A] = E[D_{a,\text{arr}}^M]. \tag{31}$$

According to Eqs. (13) and (18) we have

$$E[W_{a,q}^M] + \frac{1}{\mu_a^M} + E[W_{e,q}^M] + \frac{1}{\mu_e} = E[W_{a,q}^A] + \frac{1}{\mu_a^A}. \tag{32}$$

Since $E[W_{e,q}^M] + \frac{1}{\mu_e} \geq 0$, and substituting it into Eq. (32) yields

$$E[W_{a,q}^M] + \frac{1}{\mu_a^M} \leq E[W_{a,q}^A] + \frac{1}{\mu_a^A}. \tag{33}$$

Next, we define $f(\mu_a) = E[W_{a,q}] + \frac{1}{\mu_a}$ and expand it to

$$\begin{aligned} f(\mu_a) &= \frac{(\rho_a)^2 + (\lambda_1 + \lambda_2)^2 \sigma^2(S_a)}{2(\lambda_1 + \lambda_2)(1 - \rho_a)} + \frac{1}{\mu_a} \\ &= \frac{(k_r + 1)(\lambda_1 + \lambda_2)}{2k_r \mu_a (\mu_a - (\lambda_1 + \lambda_2))} + \frac{1}{\mu_a} \end{aligned} \tag{34}$$

It can be proved that $f(\mu_a)$ is monotonically decreasing when $\mu_a > (\lambda_1 + \lambda_2)$, which is the necessary condition for the system to reach steady states. Therefore, according to Eqs. (33) and (34), we have

$$\mu_a^A \leq \mu_a^M \tag{35}$$

which means the service rate of airport A2 tuned for arrival delay in the AND model is smaller than that of in the MATND model.

Substituting $\mu_a^A \leq \mu_a^M$ into Eqs. (17) and (25), we have

$$E[D_{a,\text{dep}}^A] \geq E[D_{a,\text{dep}}^M] \tag{36}$$

The estimated departure delay of airport A2 in AND model would be larger than the actual departure delay.

Hence, when AND and MATND are tuned to make the model estimated arrival delay of $f'$ equals to the actual value, MATND can accurately predict the departure delay of $f$ but AND overestimates the departure delay of $f$.

**Case 2.** Suppose we tune both models to make their estimated departure delay of $f$ equals to the actual value, $E[D_{a,\text{dep}}]$:

$$E[D_{a,\text{dep}}] = E[D^A_{a,\text{dep}}] = E[D^M_{a,\text{dep}}]. \tag{37}$$

Then we have $\mu^M_a = \mu^A_a$ according to Eqs. (17) and (25). Substituting it to Eq. (37) yields

$$\begin{aligned}[D^M_{a,arr}] &= E[W^M_{a,q}] + \frac{1}{\mu^M_a} + E[W^M_{e,q}] + \frac{1}{\mu_e} \\ &= E[W^A_{a,q}] + \frac{1}{\mu^A_a} + E[W^M_{e,q}] + \frac{1}{\mu_e} \\ &= [D^A_{a,arr}] + E[W^M_{e,q}] + \frac{1}{\mu_e}. \end{aligned} \tag{38}$$

Since $E[W^M_{e,q}] + \frac{1}{\mu_e} \geq 0$, we have

$$E[D^A_{a,arr}] \leq E[D^M_{a,arr}] = E[D_{a,arr}]. \tag{39}$$

Hence, when AND and MATND are tuned to make the model estimated departure delay of $f$ equals to the actual value, MATND can accurately predict the arrival delay of $f'$ well but AND underestimates the departure delay of $f$.

**Propositions 1 and 2** together show that the proposed MATND model is superior to the AND model in the sense that it is able to accurately estimate the expected departure delay and expected arrival delay, simultaneously, while AND cannot, when there is en-route congestion.

**4 Testing of MATND on a stylized network**

In this section, we demonstrate the importance of including en-route congestion in a network delay model using a stylized network. We construct a simple network with three airports (A1, A2 and A3) and an en-route point (E1) to represent en-route airspace that is likely to be congested under heavy traffic. We run a discrete event simulation on this stylized network to obtain the flight delay statistics on this network. Then we apply the classic network delay model, AND, and the proposed model, MATND, on this simple network experimenting a number of parameter tuning schemes. Lastly, we compare the average delays obtained from AND and MATND with the ones from discrete event simulation. The results show that MATND can competently model the delays on the simple network while AND cannot when the en-route airspace is congested. Hence, it is important to include en-route congestion nodes in a network delay model when en-route congestion is significant.

**4.1 Setup of a simple network and discrete event simulation**

A simple network is constructed as shown in Fig. 23, which has three airports (A1, A2 and A3) and an en-route point (E1) to represent en-route airspace that is likely to be congested. We assume unidirectional traffic departing from A1, with potential en-route congestions between A1 and A2, connecting at A2, and arriving at A3 at the end.

We run a discrete event simulation on this stylized network to obtain the flight delay statistics. In the discrete event simulation, we simulate a batch of flights departing at A1 following a Poisson distribution with an average rate of departure demand $\lambda_{dep}$, with potential delays at A1, E1, A2, and A3. The actual time of each flight to depart / arrive at A1, E1, A2, and A3 is calculated based on the service time following Erlang distributions Erlang($k_r, \mu$). A1, A2, A3 are set to have the same average service rate (15 flights per 15 min) and Erlang order $k_r = 2$. Note that there is only departure (arrival) traffic at A1 (A3), while both arrival and departure traffic are processed by a single server at A2. E1 is set to have a reduced average service rate (10 flights per 15 min) and Erlang order $k_r = 1$, to represent an en-route congestion area with limited capacity and large uncertainty in service time. The actual flight time of each flight on each sector is draw randomly from Gaussian distribution $N(FT_{avg}, \sigma)$, where $FT_{avg}$ is the average flight. More details about the simulation setup are provided in Appendix 3.

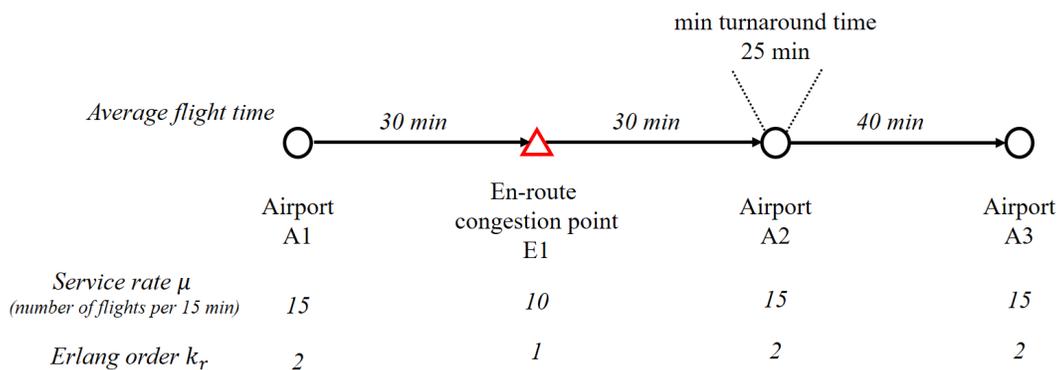

Fig. 23 The network structure of the stylized network.

To simulate different traffic conditions, we set $\lambda_{dep}$ as 6, 10, 14 flights per 15 minutes for 6 hours in each simulation cycle to represent the low, medium, and heavy traffic level, respectively. We run the simulation cycle 1000 times to obtain a representative dataset of the delays at each node of the network.

Then we use the classic network delay model, AND, and the proposed model, MATND, to estimate delays on this stylized network and compare the results with the ones from the discrete event simulation. Since both AND and MATND are network delay models whose queueing parameters of each can be adjusted, we test two parameter-tuning schemes. One scheme assumes that the queueing parameters are known; while in the second scheme, the queueing parameters are unknown and need to be estimated by searching for the values that give the best match against the simulation result.

**4.2 Testing of MATND and AND with known queueing parameters**

In this test, we assume the parameter values in the discrete-event simulation are known and use that as the queueing parameters in MATND and AND. The only difference between MATND and AND model lies in whether to have the en-route point modeled in the delay network or not. The queueing parameters are listed in Table 2.

Table 2 Specification of MATND and AND with known queueing parameters.

| Node | Simulation | | MATND | | AND | |
|---|---|---|---|---|---|---|
| | Service rate | Erlang order $k_r$ | Service rate | Erlang order $k_r$ | Service rate | Erlang order $k_r$ |
| A1 | 15 | 2 | 15 | 2 | 15 | 2 |
| A2 | 15 | 2 | 15 | 2 | 15 | 2 |
| A3 | 15 | 2 | 15 | 2 | 15 | 2 |
| E1 | 10 | 1 | 10 | 1 | - | - |

Both MATND and AND are able to generate delay estimates very quickly. For the three different traffic conditions (i.e., $\lambda_{dep} \in [6, 10, 14]$), the computational time are 1.17, 2.00, 2.45 seconds for MATND and 0.90, 1.45, 1.78 seconds for AND on a personal PC respectively, while it takes 64.25, 154 69, 229.37 seconds for the discrete-even simulation to run 1000 cycles.

The estimated average delay of the departure / arrival process at each node by MATND and AND compared with the discrete-event simulation results under low, medium, and high traffic conditions are summarized in Fig. 24. The blue boxplots show the distribution of delays from discrete-event simulation, where the red lines show the average values. The orange bars represent the expected delays calculated by MATND, and the grey bars are calculated by AND. According to Fig. 24, AND consistently underestimates the average delays, while MATND captures the average delays better. With increased traffic level, MATND outperforms AND by a larger margin.

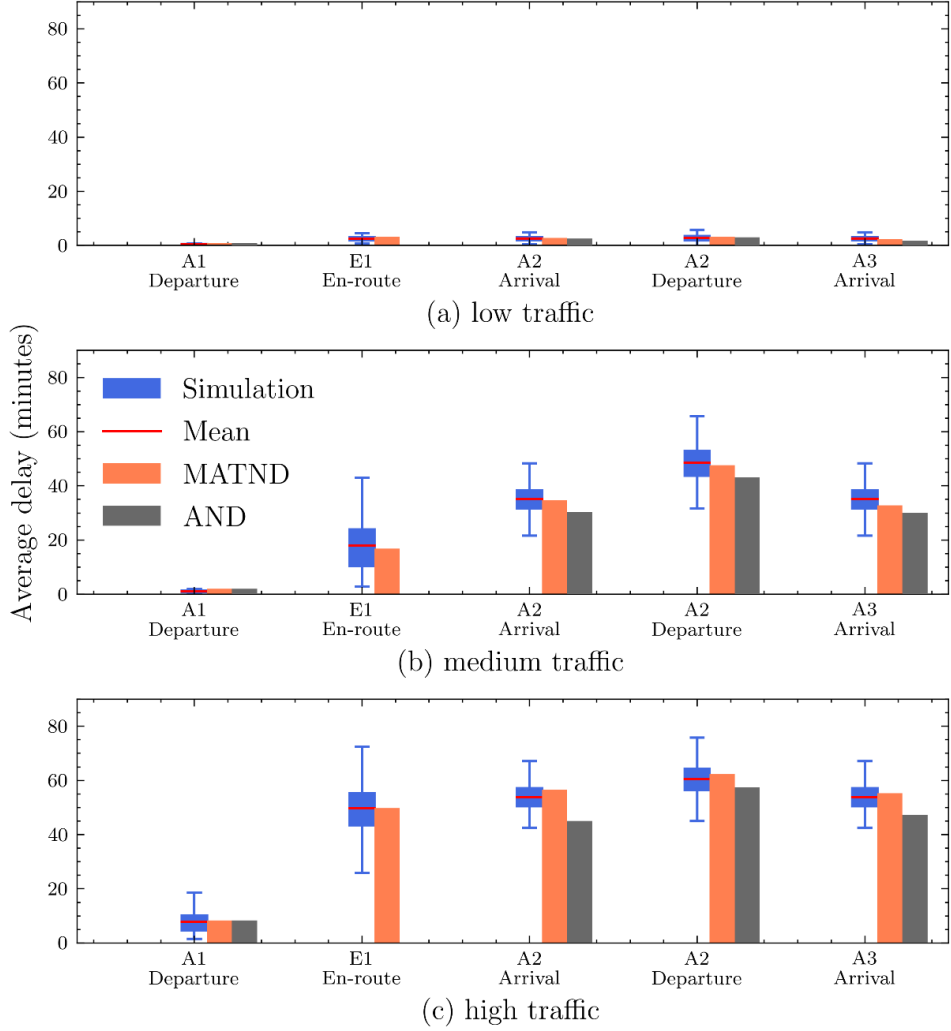

Fig. 24 Delays under different traffic levels with known queueing parameters.

**4.3 Testing of MATND and AND by tuning queueing parameters**

In this test, we assume that we have no knowledge of the parameter values used in discrete-event simulation, which is more representative of the real-world situation. We tune the queue parameters in MATND and AND separately to find the values that produce delay results that best match with the simulation results. Here, we set the Erland order as 2 for airport nodes and 1 for en-route node in MATND and AND as commonly used in airport queueing models. Then, we search for the best set of service rate values of the nodes in the delay network models. The node, A2, is a single server with two delay measures, arrival delay and departure delay. Therefore, different service rate values can be found depending on whether to tune for the arrival delay or the departure delay.

The optimized queuing parameters when tune for the arrival delay at A2 are summarized in Table 3. The delay results under low, medium, and high traffic conditions are summarized in Fig. 25. We observe that AND overestimates the average delay of A2 departure significantly under all traffic conditions, while MATND estimates the average delay of each process well.

Table 3 MATND and AND queueing parameters tuned for A2 arrival delay.

| Node | Simulation | | MATND (tuned for A2 arrival) | | AND (tuned for A2 arrival) | |
|---|---|---|---|---|---|---|
| | Service rate | Erlang order | Service rate | Erlang order | Service rate | Erlang order |
| A1 | 15 | 2 | 15 | 2 | 15 | 2 |
| A2 | 15 | 2 | 15 | 2 | 13 | 2 |
| A3 | 15 | 2 | 15 | 2 | 17 | 2 |
| E1 | 10 | 1 | 10 | 1 | - | - |

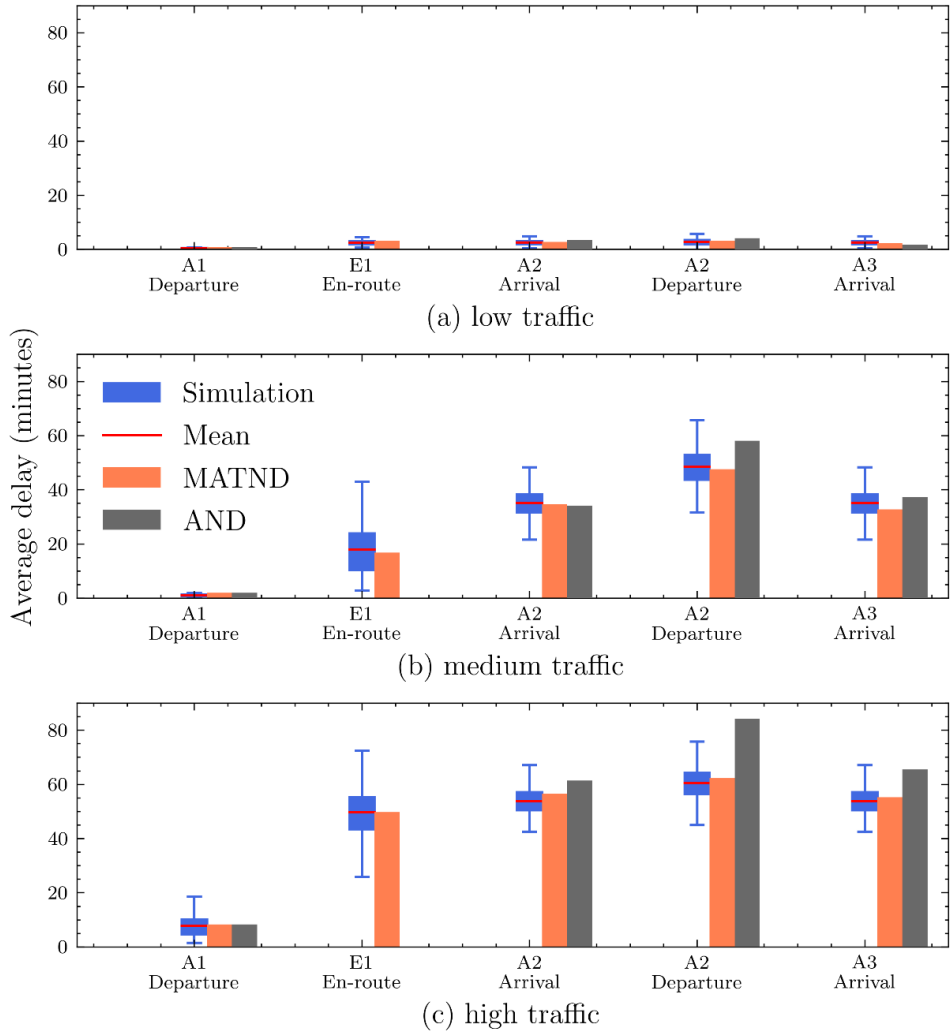

Fig. 25 Delays under different traffic levels with queueing parameters tuned for A2 arrival.

When queuing parameters are tuned to match the departure delay at A2, the parameter settings are summarized in Table 4 and the delay results are shown in Fig. 26. We observe that AND underestimates the average delay of A2 arrival significantly under all traffic conditions, while MATND estimates the average delay of each process reasonably well. Note that the queueing parameters are the same when tuned to match the arrival delay and the departure delay at A2 for MATND model, indicating MATND provides a comprehensive and accurate representation of the queueing network in this stylized example.

Table 4 MATND and AND queueing parameters tuned for A2 departure delay.

|  | Simulation |  | MATND (tuned for A2 departure) |  | AND (tuned for A2 departure) |  |
| --- | --- | --- | --- | --- | --- | --- |
| Node | Service rate | Erlang order | Service rate | Erlang order | Service rate | Erlang order |
| A1 | 15 | 2 | 15 | 2 | 15 | 2 |
| A2 | 15 | 2 | 15 | 2 | 14 | 2 |
| A3 | 15 | 2 | 15 | 2 | 16 | 2 |
| E1 | 10 | 1 | 10 | 1 | - | - |

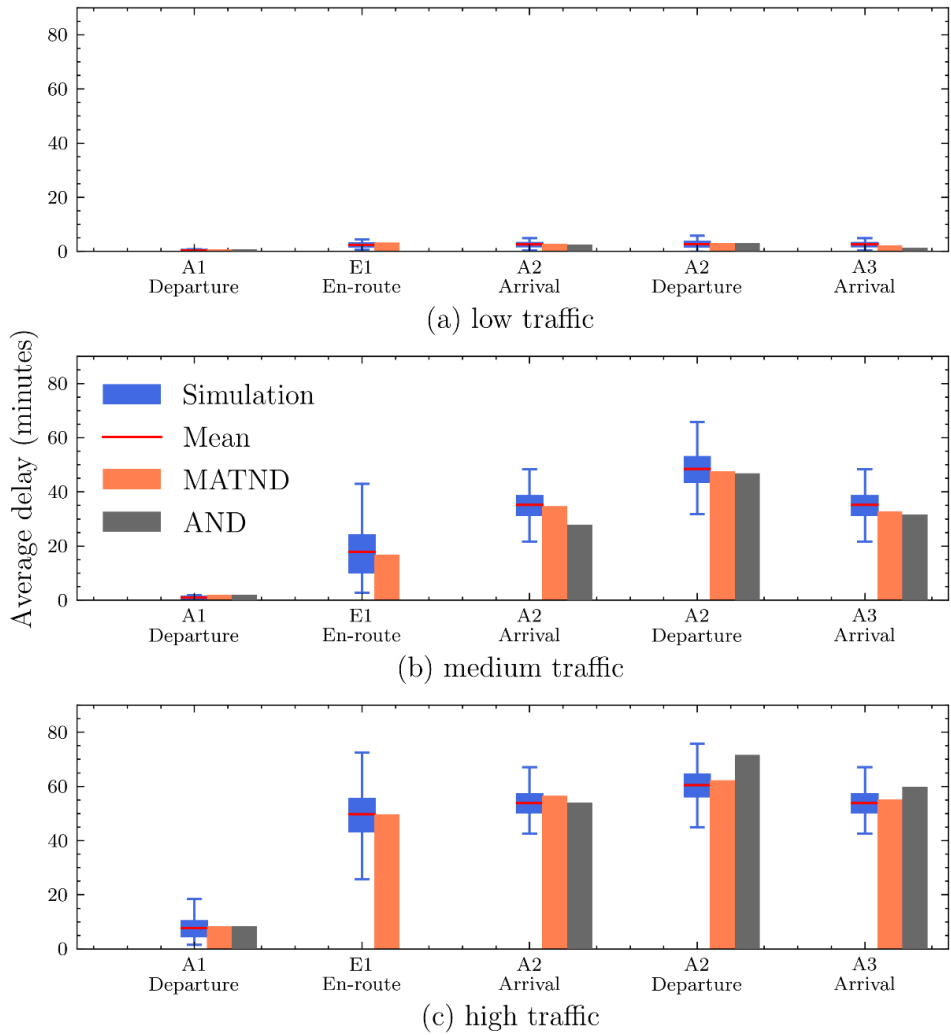

Fig. 26 Delays under different traffic levels with queueing parameters tuned for A2 departure.

To summarize, AND cannot be tuned to produce accurate estimates for both A2 arrival and departure processes at the same time. This is because the en-route congestion at E1 is not explicitly modelled in AND. To match the average delay value at A2 arrival, the service rate at A2 needs to be small, however, this would inevitably lead to an overestimation of A2 departure delay. This effect is not significant when the traffic level is low and the en-route congestion at E1 is small. However, as the traffic level increases, this over/under-estimation problem becomes more prominent.

## 5 Testing of MATND on an air traffic network of China

In this section, we demonstrate the application of the MATND model to a large-scale multi-layer air traffic network in China using a set of historical aircraft tracking data and flight schedule data over China airspace from November 1 to 30, 2016. In this dataset, we excluded OD pairs whose traffic flow averages less than one flight per day. As a result, 836 OD pairs in China remain for analysis. The total number of scheduled commercial flights over the period of interest is 261,609. Among these flights, 169,969 (65% of the total) can be captured by the aircraft tracking data collected from Flightradar24. The discrepancy is due to four reasons: 1) some aircraft are not equipped with an ADS-B transponder or a Mode S transponder; 2) the aircraft may be flying in areas with little or no Multilateration (MLAT) coverage; 3) the aircraft might be flying outside of Flightradar24's receiver coverage; and (or) 4) the aircraft may be blocked in Flightradar24's system by the owner or operator.

A multi-layer air traffic network was constructed based on these data. The network includes 1376 air routes linking the 56 busiest airports in China and 30 en-route congestion points. The parameter estimation results for the queuing system at each airport and en-route congestion point are listed in Appendix A1.

### 5.1 Model validation

To test the model performance, we use MATND to predict flight delays across China's national air traffic network and compare the results with the delays predicted by the AND model, benchmarking with the actual delays on that day.

Fig. 27 shows the average delay per departure/arrival at each of the 25 busiest airports over one month. The results predicted by MATND (orange bars) are shown together with those by AND (grey bars) and actual observations (blue bars). MATND predictions are closer to the actual value for both departure delays and arrival delays, whereas the AND predictions are much lower than the actual values. Root Mean Square Errors (RMSEs) and Mean Absolute Percentage Errors (MAPEs) of MATND for both departure and arrival delays are much smaller than the RMSEs and MAPEs of AND, as shown in Table 5.

MATND's predictions are closer to actual values than AND in both departure and arrival delays as shown in Fig. 27 because AND does not account for delays due to en-route congestion, while MATND does account for en-route congestions. The impact of en-route congestion is significant for the air traffic network of China, where the airspace resources are limited.

Both MATND and AND underestimate departure delays in comparison with arrival delays. This is due to the effect of Air Traffic Flow Management, i.e., Ground Delay Programs (GDPs), or ATC coordination mechanisms similar to GDP. Departure flights are asked to wait on the ground, rather than hold in the air, when the destination airport is likely to be congested. Hence, significant departure delays are generated on the ground for the departure flights, and it is not explicitly modelled in MATND or AND.

Neither MATND or AND can fully capture the amount of actual delays from operational data. This is because MATND and AND do not model airline reactions to congestion and does not capture other sources of delay, such as aircraft mechanical problems. The value of MATND, AND, and this kind of analytical model based on queuing network, is to "re-produce trends and behaviors that are observed in practice in the NAS system, not whether it can reproduce exactly the actual delays measured," as stated in (Pyrgiotis et al., 2013). The critical test is whether the model produces estimations of delays due to local capacity constraint and network propagations are of the same order of magnitude as delays measured in practice. Results shown in Fig. 27, Fig. 28 and Fig. 29 confirms that both MATND and AND can capture the trends of delays across the network.

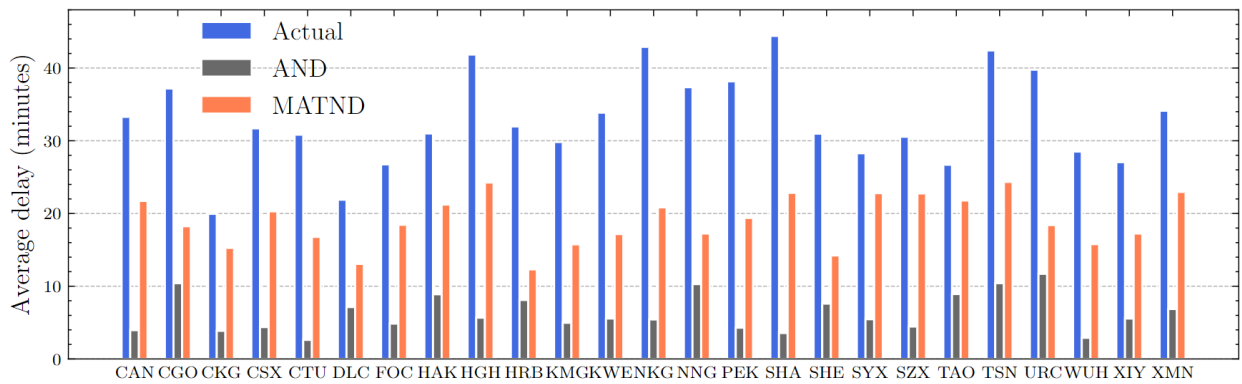

a) Departure delay

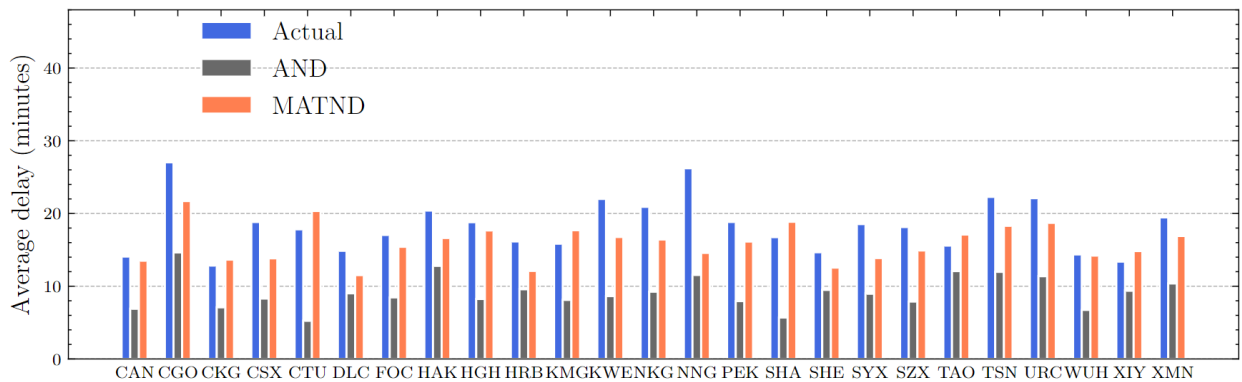

b) Arrival delay

Fig. 27 Flight delay prediction at 25 busiest airports.

Table 5 Delay prediction statistics of MATND and AND.

|  | **MATND** | | **AND** | |
| --- | --- | --- | --- | --- |
|  | RMSE (minutes) | MAPE (%) | RMSE (minutes) | MAPE (%) |
| **Departure delay** | 14.87 | 40.95 | 27.24 | 80.42 |
| **Arrival delay** | 3.90 | 16.24 | 9.53 | 49.08 |

Fig. 28 and Fig. 29 present an hourly summary of flight operations at Guangzhou Baiyun International Airport (CAN) on a sample day. Fig. 28 shows the number of departures/arrivals at CAN predicted by MATND (orange line), AND (grey line), as well as the actual values (blue line). Fig. 28 and Fig. 29 show an example of how the MATND model generates expected delays over fixed time intervals at an airport, capturing the trend of delays over a day. However, the magnitude of the delays may not exactly match the operational data on a particular day at an airport due to the stochastic nature of the delay process. When there are only a few flights in a time interval, the estimated delay amount could be significantly different from the actual value. This is why there are spikes and dips around midnight in the departure delay predictions by MATND. For example, the spike at hour 24 in Fig. 29 (a) is caused by a flight with a significant delay estimation by MATND. In the actual flight data, there were four flights departed between Hour 24 and Hour 1. Their delays are 22 min, 28 min, 18 min, and 28 min, and the average is 24 min. In MATND, there were two flights departed in the same time window. One flight has a departure delay of 9 min, another one has a departure delay of 83 min, and the average of the two flights is 46 min. The delay of 83 min is caused by en-route congestion rather than airport delay.

Plots similar to Fig. 28 and Fig. 29 over 30 days at CAN are shown in Appendix A4 to give a more comprehensive view of the model performance. Based on these figures, we observe that both AND and MATND provide accurate airport throughput predictions. MATND can capture the trends of departure delays and arrival delays, under-estimates departure delay (mainly due to GDPs), and performs better than AND on both departure and arrival delay in most cases.

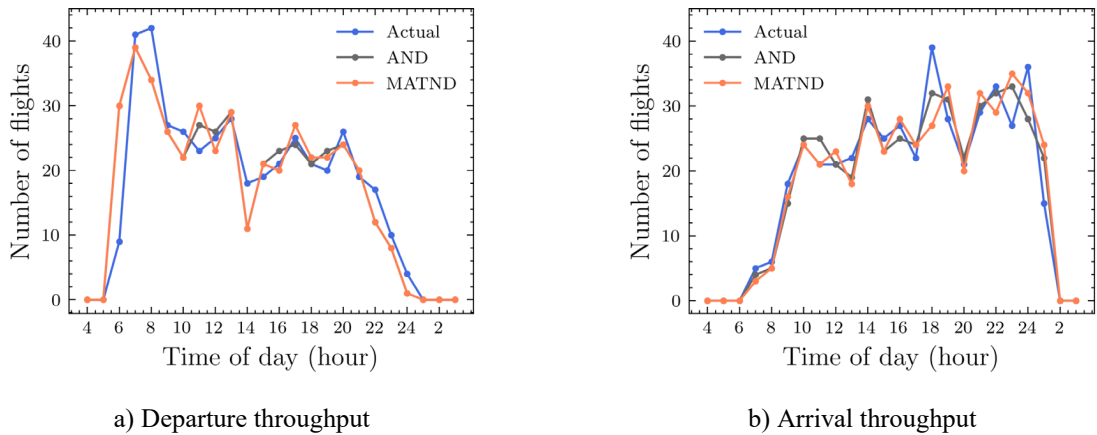

a) Departure throughput  b) Arrival throughput

Fig. 28 Throughput prediction at CAN by hour of the day.

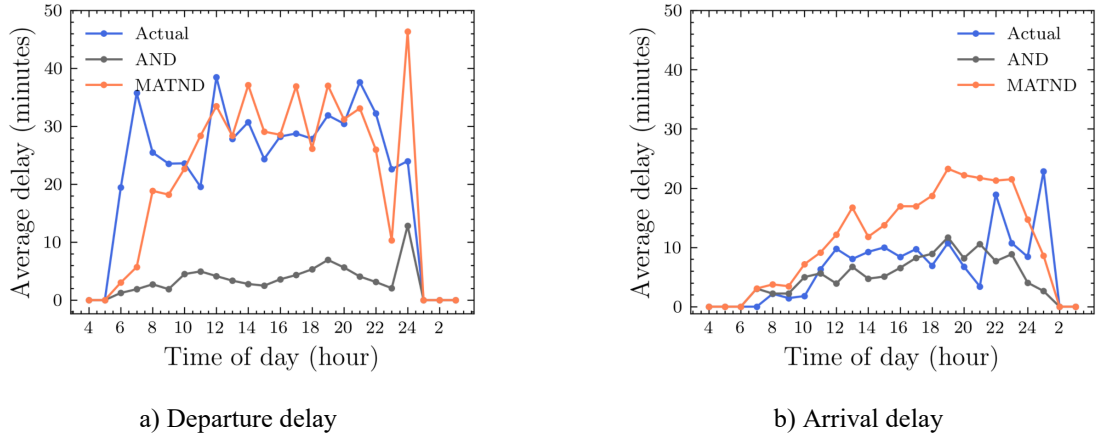

Fig. 29 Delay prediction at CAN by hour of the day.

## 5.2 Delay analysis of China air traffic network

MATND can be used to understand bottlenecks in the current network regarding traffic congestion. Using MATND, we can trace the flight delay over a network by whether it is caused by an airport or en-route local congestion or network propagation effect. The detailed calculation method of how much delay is caused by a local (either airport or en-route congestion) source and how much is due to upstream sources is described in Appendix A2.

A summary of flight delays caused by different sources over China's air traffic network based on the operations in November 2016 is calculated and shown in Table 6. The average delay experienced by a flight is ~16 min, which is consistent with the actual situation in China in 2016 (CAAC, 2016, 2018). The local delay incurred at en-route congestion points (5.0 min) is smaller than the local delay at airports (6.4 min) on average. However, the propagated delay at en-route congestion points (15.6 min) is greater than the propagated delay at airports (9.5 min) on average. Fig. 30 shows the expected local delay and propagated delay at the 25 busiest airports in China. These results show that delay propagation effects are more severe at en-route congestion points than at airports.

Table 6 Summary of flight delays caused by different sources over China's air traffic network based on the operations in November 2016.

| (minutes per flight) | At airports | At en-route congestion points |
| --- | --- | --- |
| Average local delay | 6.43 | 4.99 |
| Average propagated delay | 9.54 | 15.57 |

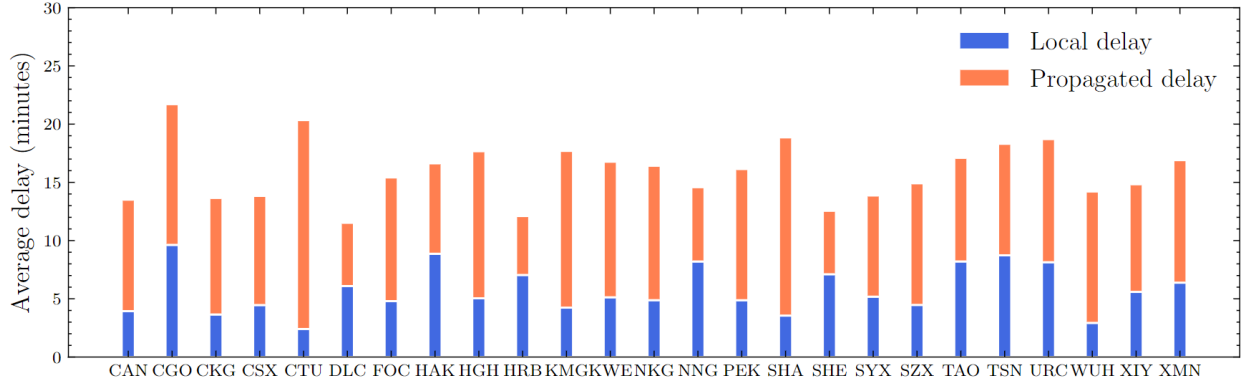

Fig. 30 Expected flight delay per flight at 25 busiest airports.

## 6 Demonstration of MATND for evaluating improvement strategies

In this section, we demonstrate how to use MATND for evaluating air traffic network improvement strategies. Using MATND, we can quickly conduct "what-if" scenario analysis to estimate the effectiveness of infrastructure investment, technology improvement, or managerial changes on delay reduction at the system-level for the current network. For example, the following questions can be answered by running simulated scenarios via MATND:

- With a limited airport expansion budget, which airports in the national network should be prioritized to invest if the goal is to reduce local flight delay and overall network delay?

- If airport expansion is not possible in the near term due to financial or policy constraint, how much delay can be reduced by improving en-route capacity?

- Are there any airports that would benefit more from en-route capacity improvement rather than runway expansion?

### 6.1 Airport capacity improvement

In this analysis, the improvement strategy is to increase the capacity of the airport runway. We use MATND to test how much delay would be reduced if an airport is expanded by building one more independent runway. The number of runways and the physical layout of the runway system are two important factors for determining the airport capacity (Jacquillat and Odoni, 2018). Most airports in China have parallel runways which meet the requirement for independent operations. In this analysis, we assume all airports in this network have independent parallel runways, and adding a new parallel runway would increase the service rate of the queuing system at the airport increased by $1/n$ times, where $n$ is the number of existing runways at this airport. For example, there are currently two runways at Chengdu Shuangliu International Airport (CTU) and the capacity of this airport is 52 flights per hour. Adding one new independent runway would increase the capacity to 78 flights per hour.

Fig. 31 shows the reduction in flight delays across the whole network when one new runway is built at each of the 25 busiest airports. The results indicate that the impact of adding a new runway is dependent on the airport, and the delay reduction is very limited if investing in only one airport. On average, the total delay reduction is 0.185 min per flight across the whole network. The maximum delay reduction, 0.307 min per flight, occurs when a new runway is built at PEK. The minimum delay reduction is 0.027 min per flight at Wuhan Tianhe International Airport (WUH).

Moreover, building a new runway may even increase network propagation delay. The blue bar and orange bar represent the reduction minutes of local delay and propagated delay, respectively. We can see that there are orange bars with negative values in Fig. 31, which indicates that building a new runway at these airports (CAN, CKG, SZX, and WUH) can increase network delay propagation. This result suggests that releasing more quickly of aircraft from some airports could aggravate en-route congestion levels, which in turn increases the propagated delay within the whole network.

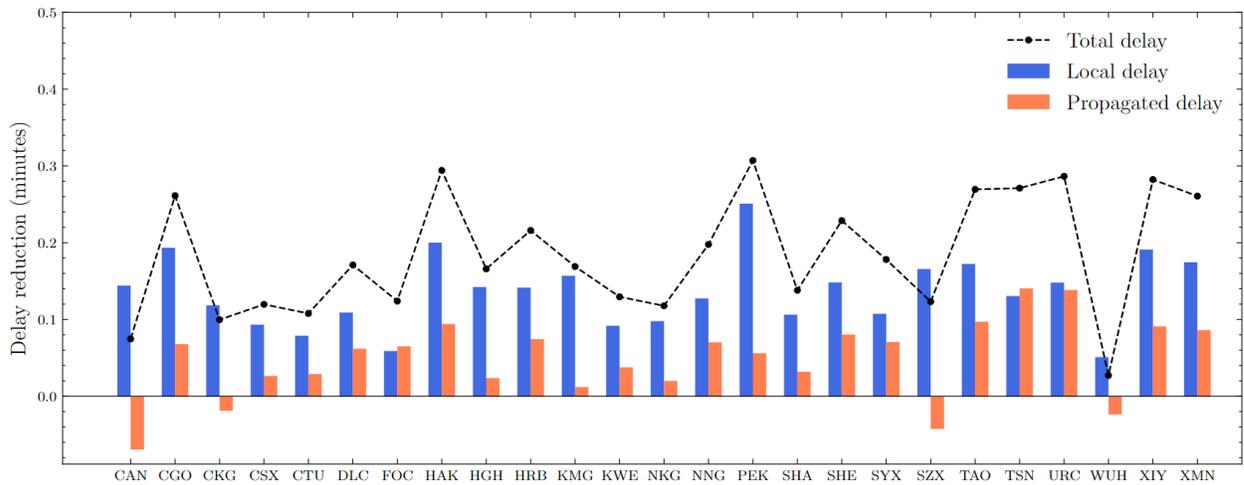

Fig. 31 Delay reduction per flight over the network from building a new runway at the airport.

We can further evaluate how many airports need to be expanded to have a significant delay reduction over the whole network via MATND. We rank airports by the resultant system-level delay reduction if a new runway is built at each airport, then simulate how much delay reduction can be achieved if these runways are added into the system accumulatively. Fig. 32 shows that a 20% delay reduction can be achieved at the system-level if all top 10 airports are expanded with a new runway. The marginal reduction in delay decreases with the number of airports joining this program increases.

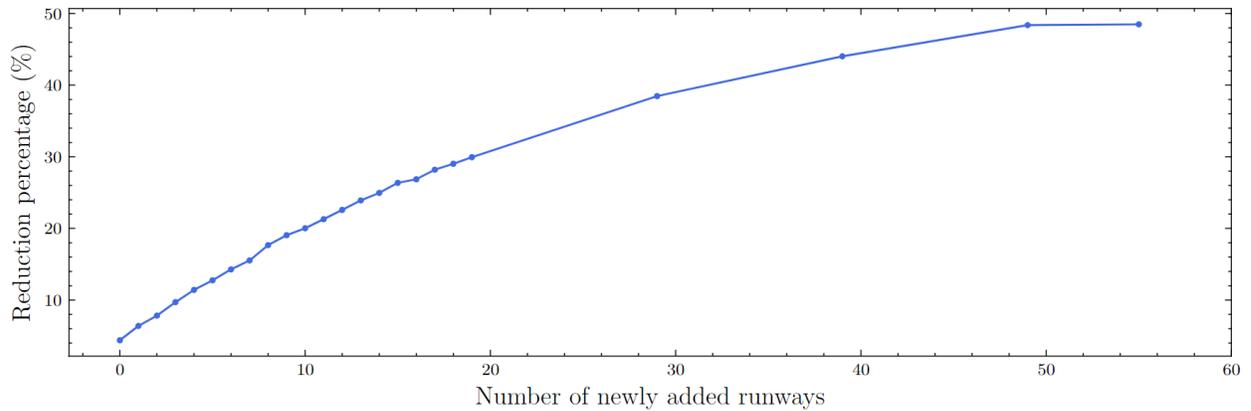

Fig. 32 System-level delay reduction by the number of airports to be expanded.

**6.2 En-route capacity improvement**

Airport expansion, especially building new runways, is capital intensive and subjected to many social-political constraint. In contrast, increasing airspace capacity could be easier and faster to be implemented. Airspace capacity may be realized by opening up more airspace for civil aviation activities, modifying operational procedures or rules (e.g., re-design sector, more progressive traffic management), deploying better surveillance and communication technologies, or developing advanced operational concepts (e.g., dynamic airspace and 4D trajectories). In this analysis, we computed the amount of delay reduction that can be achieved if the capacity of en-route congestion points increases hypothetically by 10% to 200%.

As shown in Fig. 33, if the en-route capacity is improved by just 20%, a 32% delay reduction in total flight delays can already be reached, while the greatest delay reduction is 41% when the en-route capacity is doubled. It should also be noted that the delay reduction from en-route capacity improvement is capped at around 40%, where the continued improvement of 60% - 200% extra en-route capacity has a negligible effect.

Fig. 34 shows how much delay reduction at the 25 busiest airports can be achieved if the capacity of en-route congestion points increases by 20%. The local delays would remain almost the same, while the propagated delays would reduce significantly. Moreover, the top 5 airports with largest delay reduction from en-route capacity improvement are CTU, SHA, HGH, PEK and KMG. This indicates that the throughput of these airports are constrained by en-route capacity significantly.

In summary, this analysis demonstrated that the impact of en-route capacity improvement could be significant. A small increase in en-route capacity could lead to a substantial reduction in network flight delay in the air traffic network of China where limited airspace is open for civil aviation.

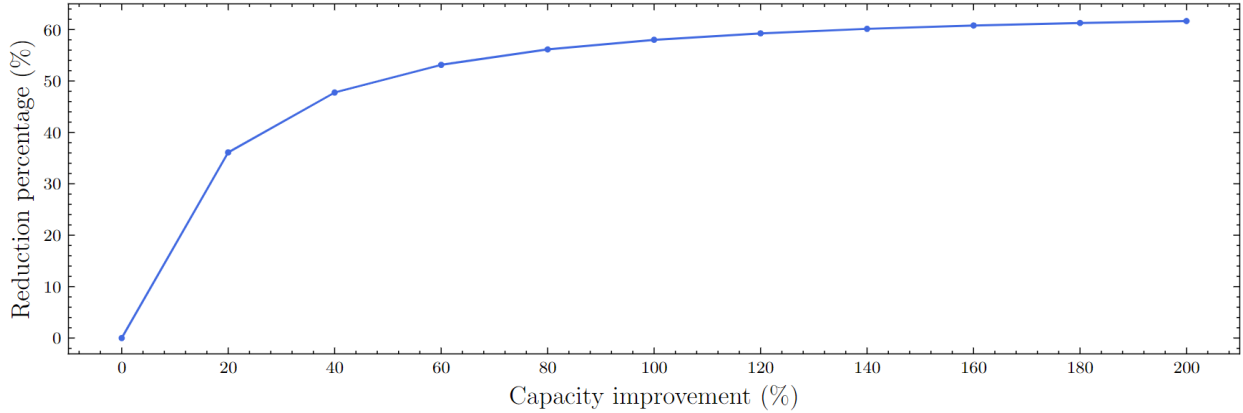

Fig. 33 System-level delay reduction by the capacity improvement of en-route congestion points.

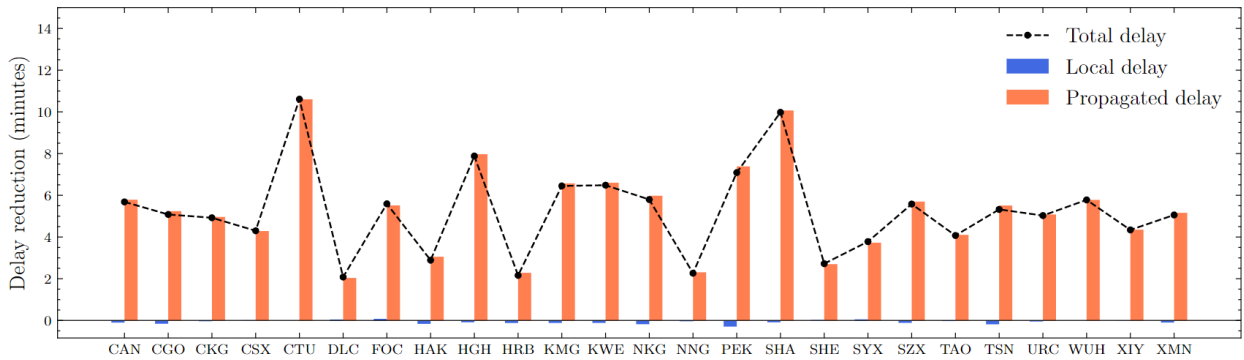

Fig. 34 Flight delay changes if en-route capacity improves by 20%.

**6.3 Individual airports' response to en-route and runway capacity improvement**

In this analysis, we compare the effectiveness of en-route capacity improvement and runway capacity improvement for individual airports, focusing on delay reduction of flights to/from that airport. We identify if any airports that would benefit more from mitigating en-route congestion than enhancing runway capacities, and vice versa. For each airport, we use MATND to compute potential delay reductions that can be achieved for flights to/from that airport if a new runway is built or if the en-route congestion connected to that airport is eliminated.

Counter-intuitively, 14 out of the 25 airports being analyzed would benefit more from en-route capacity improvement rather than runway capacity improvement, as shown in Fig. 35. The prominent ones in this category are Chengdu airport (CTU), Shanghai Hongqiao Airport (SHA), and Kuming Changhui Airport (KMG). The other 11 airports will indeed reduce more delay if a runway is built than en-route capacity increases. The results confirm that the capacity bottleneck in the current air traffic network of China is en-route airspace constraint rather than runway capacity for more than half of the major airports.

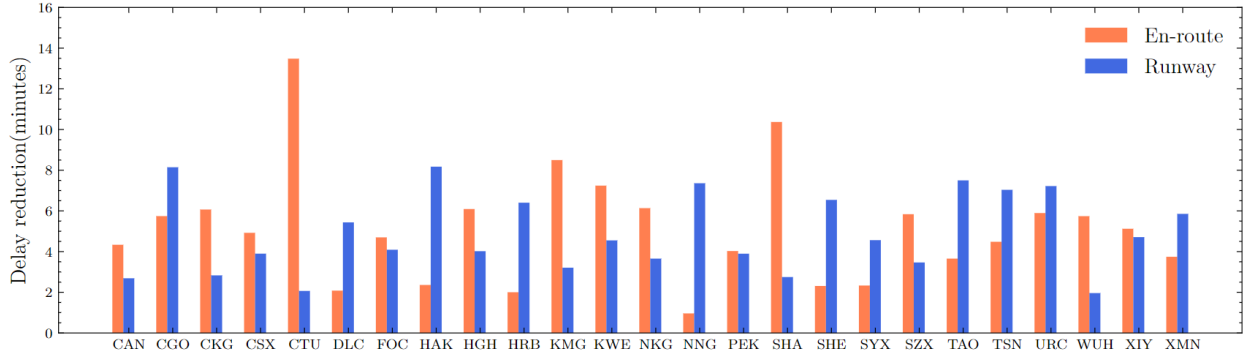

Fig. 35 Comparison of delay reduction at 25 busiest airports due to en-route capacity improvement and runway capacity improvement.

## 6.4 Discussion

In summary, the results indicate that the airspace constrains introduced significant delays in China's current aviation system. Although expanding ground aviation infrastructure (building new airports or expending existing airports) could reduce some amount of flight delay on average, however, projects focusing on en-route capacity improvement can bring immediate reduction on delays without a significant amount of financial investment. For a few airports, like Chengdu airport (CTU), Shanghai Hongqiao Airport (SHA), and Kuming Changhui Airport (KMG), the delay reduction would be larger from en-route capacity improvement than from runway capacity expansion.

These findings are consistent with other reports and studies. Dong and Ryerson (2019) summarized that there is a mismatch between the rapidly expanding ground aviation infrastructure and a highly constrained airspace in China's aviation system based on an examination of academic literature, official reports, and news articles. The recent CAAC reports confirms that Air Traffic Management Bureau (ATMB) is making efforts along the direction to improve en-route air traffic capacity. ATMB identifies the choke points in the current system and transforms bidirectional routes into unidirectional air corridors to alleviate the operational pressure and enhance flight efficiency. Untill 2018, five corridors were established to replace the single trunk route serving big terminal manoeuvring areas (TMAs) (CAAC, 2019).

## 7 Conclusion

We have developed a novel data-driven method for air transport network abstraction and simulation, named as Multi-layer Air Traffic Network Delay (MATND) model. In this model, en-route constraint are explicitly modeled within a multi-layer air traffic network consisting of airports, en-route congestion points, and the air routes linking them. Then, it computes delays due to congestion at individual airports and en-route congestion points and estimates the propagation of delays throughout the network.

Using actual flight data, we demonstrated that MATND is a powerful tool, from a macroscopic perspective, for delay prediction and bottleneck identification in a national air transport system. We also showed that MATND can be used to evaluate a broad range of alternative policies and strategies in air traffic management improvement, such as infrastructure improvements and network structure modifications. For the current air traffic network of China, there are more than half of airports that would benefit more from en-route capacity improvement rather than runway capacity improvement for flight delay reduction.

One direction of future work is to consider Ground Delay Programs (GDPs) in the delay propagation algorithm. GDPs are an explicit or implicit part of the air traffic flow management in many countries, including China. Some airborne delays are shifted to ground delays, and hence the delay mechanism has been changed. Thus, the propagation of delays through the network needs to be modified in the MATND model. Another direction is to develop simulation-based optimization models to reduce flight delays through infrastructure improvement, better scheduling, or Air Traffic Flow Management policies.

**Acknowledgments**

The work was supported by the Hong Kong Research Grant Council General Research Fund Grant (Project No. 11209717). The authors would like to thank Dr. Jianxiang Huang for his help in graphics design and production for this paper.

**References**

AhmadBeygi, S., Cohn, A., Guan, Y., Belobaba, P., 2008. Analysis of the potential for delay propagation in passenger airline networks. J. Air Transp. Manag. 14, 221–236. https://doi.org/10.1016/j.jairtraman.2008.04.010.

Andrienko, G., Andrienko, N., Fuchs, G., Garcia, J.M.C., 2017. Clustering Trajectories by Relevant Parts for Air Traffic Analysis. IEEE Trans. Vis. Comput. Graph. 24, 34–44. https://doi.org/10.1109/TVCG.2017.2744322.

Beatty, R., Hsu, R., Berry, L., Rome, J., 1999. Preliminary Evaluation of Flight Delay Propagation through an Airline Schedule. Air Traffic Control Q. 7, 259–270. https://doi.org/10.2514/atcq.7.4.259.

Bertsimas, D., Lulli, G., Odoni, A., 2011. An integer optimization approach to large-scale air traffic flow management. Oper. Res. 59, 211–227. https://doi.org/10.1287/opre.1100.0899.

Bertsimas, D., Lulli, G., Odoni, A., 2008. The air traffic flow management problem: An integer optimization approach, in: Lecture Notes in Computer Science (Including Subseries Lecture Notes in Artificial Intelligence and Lecture Notes in Bioinformatics). Springer, Berlin, Heidelberg, pp. 34–46. https://doi.org/10.1007/978-3-540-68891-4_3.

Bertsimas, D., Patterson, S.S., 1998. The Air Traffic Flow Management Problem with Enroute Capacities. Oper. Res. 46, 406–422. https://doi.org/10.1287/opre.46.3.406.

Bilimoria, K.D., Sridhar, B., Grabbe, S.R., Chatterji, G.B., Sheth, K.S., 2001. FACET: Future ATM Concepts Evaluation Tool. Air Traffic Control Q. 9, 1–20.

https://doi.org/10.2514/atcq.9.1.1.

Bolić, T., Castelli, L., Corolli, L., Rigonat, D., 2017. Reducing ATFM delays through strategic flight planning. Transp. Res. Part E Logist. Transp. Rev. 98, 42–59. https://doi.org/10.1016/j.tre.2016.12.001.

Buchin, K., Buchin, M., Duran, D., Fasy, B.T., Jacobs, R., Sacristan, V., Silveira, R.I., Staals, F., Wenk, C., 2017. Clustering Trajectories for Map Construction, in: Proceedings of the 25th ACM SIGSPATIAL International Conference on Advances in Geographic Information Systems. ACM, New York, NY, USA, pp. 1–10. https://doi.org/10.1145/3139958.3139964.

Buschmann, S., Trapp, M., Döllner, J., 2016. Animated visualization of spatial–temporal trajectory data for air-traffic analysis. Vis. Comput. 32, 371–381. https://doi.org/10.1007/s00371-015-1185-9.

Campanelli, B., Fleurquin, P., Arranz, A., Etxebarria, I., Ciruelos, C., Eguíluz, V.M., Ramasco, J.J., 2016. Comparing the modeling of delay propagation in the US and European air traffic networks. J. Air Transp. Manag. 56, 12–18. https://doi.org/10.1016/j.jairtraman.2016.03.017.

CAPA, 2018. European airspace control. The promise: delays. The need: action [WWW Document]. URL https://centreforaviation.com/analysis/reports/european-airspace-control-the-promise-delays-the-need-action-424075 (acessed 6.4.20).

Chen, Z., Shen, H.T., Zhou, X., 2011. Discovering popular routes from trajectories, in: 2011 IEEE 27th International Conference on Data Engineering. IEEE, pp. 900–911. https://doi.org/10.1109/ICDE.2011.5767890.

Cho, J.Y., Welch, J.D., Underhill, N.K., 2011. Analytical workload model for estimating en route sector capacity in convective weather, in: 9th USA/Europe ATM R&D Seminar. Berlin, Germany.

Civil Aviation Administration of China (CAAC), 2019. Facts on Civil Aviation Airspace of China in 2018 [WWW Document] URL http://www.atmb.net.cn/UploadFiles/20190704085843234.pdf (accessed 6.4.20)..

Civil Aviation Administration of China (CAAC), 2018. 2017 National Civil Aviation Flight Operational Efficiency Report [WWW Document] URL http://www.caac.gov.cn/XWZX/MHYW/201803/P020180329429997641224.pdf. (accessed 6.4.20).

Civil Aviation Administration of China (CAAC), 2016. 2015 National civil aviation flight operational efficiency report [WWW Document] URL http://www.cata.org.cn/HYYJ/HYYJ/201607/P020160722436820051914.pdf (accessed 6.4.20).

Conde Rocha Murca, M., DeLaura, R., Hansman, R.J., Jordan, R., Reynolds, T., Balakrishnan, H., 2016. Trajectory clustering and classification for characterization of air traffic flows, in: 16th AIAA Aviation Technology, Integration, and Operations Conference. American Institute of Aeronautics and Astronautics, Reston, Virginia. https://doi.org/10.2514/6.2016-3760.

Delahaye, D., Puechmorel, S., 2000. Air traffic complexity: towards intrinsic metrics, in:

Proceedings of the Third USA/Europe Air Traffic Management R & D Seminar.

Dong, X., Ryerson, M.S., 2019. Increasing civil aviation capacity in China requires harmonizing the physical and human components of capacity: A review and investigation. Transp. Res. Interdiscip. Perspect. 1, 100005. https://doi.org/10.1016/j.trip.2019.100005.

Eckstein, A., 2009. Automated flight track taxonomy for measuring benefits from performance based navigation, in: Proceedings of the 2009 Integrated Communications, Navigation and Surveillance Conference, ICNS 2009. https://doi.org/10.1109/ICNSURV.2009.5172835.

Enriquez, M., 2013. Identifying temporally persistent flows in the terminal airspace via spectral clustering, in: Tenth USA/Europe Air Traffic Management Research and Development Seminar (ATM2013)/Federal Aviation Administration (FAA) and EUROCONTROL. Chicago, IL, USA, pp. 10–13.

Ester, M., Kriegel, H.-P., Sander, J., Xu, X., 1996. A density-based algorithm for discovering clusters in large spatial databases with noise, in: Proceedings of the 2nd International Conference on Knowledge Discovery and Data Mining. AAAI Press, pp. 226–231.

Eurocontrol, 2020. ESCAPE: world-class ATC real-time simulator [WWW Document]. URL https://simulations.eurocontrol.int/solutions/escape-world-class-atc-real-time-simulator (accessed 6.4.20).

Eurocontrol, 2010. Network operations report for 2010: Monitoring & reporting [WWW Document] URL https://www.eurocontrol.int/sites/default/files/publication/files/network-oper-15 ations-report-14 2010-yearly.pdf (accessed 6.4.20).

Federal Aviation Administration (FAA), 2019. NextGen implementation plan 2018-19 [WWW Document] URL https://www.faa.gov/nextgen/media/NextGen_Implementation_Plan-2018-19.pdf (accessed 6.16.20).

Ferreira, N., Klosowski, J.T., Scheidegger, C.E., Silva, C.T., 2013. Vector Field k-Means: Clustering Trajectories by Fitting Multiple Vector Fields. Comput. Graph. Forum 32, 201–210. https://doi.org/10.1111/cgf.12107.

Fleurquin, P., Ramasco, J.J., Eguiluz, V.M., 2013. Systemic delay propagation in the US airport network. Sci. Rep. 3, 1159. https://doi.org/10.1038/srep01159.

Fleurquin, P., Ramasco, J.J., Eguíluz, V.M., 2014. Characterization of delay propagation in the US air-transportation network. Transp. J. 53, 330–344.

Fricke, H., Schultz, M., 2009. Delay impacts onto turnaround performance: Optimal time buffering for minimizing delay propagation, in: ATM Seminar.

Gariel, M., Srivastava, A.N., Feron, E., 2011. Trajectory Clustering and an Application to Airspace Monitoring. IEEE Trans. Intell. Transp. Syst. https://doi.org/10.1109/TITS.2011.2160628.

Hilburn, B., 2004. Cognitive complexity in air traffic control: A literature review. EEC note, 4(04), 1–80.

Hsu, K., 2014. China's airspace management challenge. U.S.-China economic and security review commission staff report [WWW Document]. URL

https://www.uscc.gov/sites/default/files/Research/China's Airspace Management Challenge.pdf (accessed 6.4.20).

Jacquillat, A., Odoni, A.R., 2018. A roadmap toward airport demand and capacity management. Transp. Res. Part A Policy Pract. 114, 168–185. https://doi.org/10.1016/j.tra.2017.09.027.

Jacquillat, A., Odoni, A.R., 2015. An integrated scheduling and operations approach to airport congestion mitigation. Oper. Res. 63, 1390–1410. https://doi.org/10.1287/opre.2015.1428.

Kafle, N., Zou, B., 2016. Modeling flight delay propagation: A new analytical-econometric approach. Transp. Res. Part B Methodol. 93, 520–542. https://doi.org/10.1016/j.trb.2016.08.012.

Khintchine, A.Y., 1932. Mathematical theory of a stationary queue. Mat. Sb. 39, 73–84.

Kim, J., Mahmassani, H.S., 2015. Spatial and Temporal Characterization of Travel Patterns in a Traffic Network Using Vehicle Trajectories. Transp. Res. Procedia 9, 164–184. https://doi.org/10.1016/j.trpro.2015.07.010.

Kivestu, P.A., 1976. Alternative Methods of Investigating the Time Depedent M/G/k queue. M.S. thesis. Massachusetts Institute of Technology.

Larsen, R.C., Odoni, A.R., 1981. Urban operation research: logistical and transportation planning methods. Prentice Hall.

Long, D., Hasan, S., 2009. Improved predictions of flight delays using LMINET2 system-wide simulation model, in: 9th AIAA Aviation Technology, Integration, and Operations Conference (ATIO), Aviation Technology, Integration, and Operations (ATIO) Conferences. American Institute of Aeronautics and Astronautics. https://doi.org/doi:10.2514/6.2009-6961.

Long, D., Lee, D., Johnson, J., Gaier, E., Kostiuk, P., 1999. Modeling air traffic management technologies with a queuing network model of the national airspace system.

Lulli, G., Odoni, A., 2007. The European air traffic flow management problem. Transp. Sci. 41, 431–443. https://doi.org/10.1287/trsc.1070.0214.

Majumdar, A., Polak, J., 2001. Estimating capacity of Europe's airspace using a simulation model of air traffic controller workload. Transp. Res. Rec. J. Transp. Res. Board 1744, 30–43. https://doi.org/10.3141/1744-05.

Malone, K.M., 1995. Dynamic queueing systems: behavior and approximations for individual queues and for networks. Massachusetts Institute of Technology.

Marla, L., Vaaben, B., Barnhart, C., 2017. Integrated disruption management and flight planning to trade off delays and fuel burn. Transp. Sci. 51, 88–111. https://doi.org/10.1287/trsc.2015.0609.

Meyn, L., Windhorst, R., Roth, K., Van Drei, D., Kubat, G., Manikonda, V., Roney, S., Hunter, G., Huang, A., Couluris, G., 2006. Build 4 of the Airspace Concept Evaluation System, in: AIAA Modeling and Simulation Technologies Conference and Exhibit. American Institute of Aeronautics and Astronautics, Reston, Virigina. https://doi.org/10.2514/6.2006-6110.

Mogford, R.H., Guttman, J.A., Morrow, S.L., Kopardekar, P., 1995. The Complexity Construct in Air Traffic Control: A Review and Synthesis of the Literature.

Murça, M.C.R., Hansman, R.J., 2018. Identification, Characterization, and Prediction of Traffic Flow Patterns in Multi-Airport Systems. IEEE Trans. Intell. Transp. Syst. 20, 1683–1696. https://doi.org/10.1109/TITS.2018.2833452.

Murça, M.C.R., Hansman, R.J., Li, L., Ren, P., 2018. Flight trajectory data analytics for characterization of air traffic flows: A comparative analysis of terminal area operations between New York, Hong Kong and Sao Paulo. Transp. Res. Part C Emerg. Technol. 97, 324–347. https://doi.org/10.1016/j.trc.2018.10.021.

Palma, A.T., Bogorny, V., Kuijpers, B., Alvares, L.O., 2008. A clustering-based approach for discovering interesting places in trajectories, in: Proceedings of the 2008 ACM Symposium on Applied Computing - SAC '08. ACM Press, New York, New York, USA, p. 863. https://doi.org/10.1145/1363686.1363886.

Pelekis, N., Kopanakis, I., Kotsifakos, E., Frentzos, E., Theodoridis, Y., 2009. Clustering Trajectories of Moving Objects in an Uncertain World, in: 2009 Ninth IEEE International Conference on Data Mining. IEEE, pp. 417–427. https://doi.org/10.1109/ICDM.2009.57.

Peterson, Michael D., Bertsimas, D.J., Odoni, A., 1995. Decomposition Algorithms for Analyzing Transient Phenomena in Multiclass Queueing Networks in Air Transportation. Oper. Res. 43, 995–1011.

Peterson, Michael D, Bertsimas, D.J., Odoni, A.R., 1995. Models and algorithms for transient queueing congestion at airports. Manage. Sci. 41, 1279--1295.

Pollaczek, F., 1930. Über eine Aufgabe der Wahrscheinlichkeitstheorie. I. Math. Zeitschrift 32, 64–100. https://doi.org/10.1007/BF01194620.

Prandini, M., Piroddi, L., Puechmorel, S., Brazdilova, S.L., 2011. Toward Air Traffic Complexity Assessment in New Generation Air Traffic Management Systems. IEEE Trans. Intell. Transp. Syst. 12, 809–818. https://doi.org/10.1109/TITS.2011.2113175.

Pyrgiotis, N., Malone, K.M., Odoni, A., 2013. Modelling delay propagation within an airport network. Transp. Res. Part C Emerg. Technol. 27, 60–75. https://doi.org/10.1016/j.trc.2011.05.017.

Rebollo, J.J., Balakrishnan, H., 2014. Characterization and prediction of air traffic delays. Transp. Res. Part C Emerg. Technol. 44, 231–241. https://doi.org/10.1016/j.trc.2014.04.007.

Rehm, F., 2010. Clustering of flight tracks, in: AIAA Infotech@Aerospace 2010. American Institute of Aeronautics and Astronautics, Reston, Virigina. https://doi.org/10.2514/6.2010-3412.

Ren, P., Li, L., 2018. Characterizing air traffic networks via large-scale aircraft tracking data: A comparison between China and the US networks. J. Air Transp. Manag. 67, 181–196. https://doi.org/10.1016/j.jairtraman.2017.12.005.

Russo, R., 2016. Capacity planning and assessment network operations planning [WWW Document]. URL https://www.icao.int/ESAF/Documents/meetings/2016/Air Traffic

Services System Capacity 2016/1 Network Operations Planning.pdf (accessed 6.4.20).

Salaun, E., Gariel, M., Vela, A.E., Feron, E., 2012. Aircraft Proximity Maps Based on Data-Driven Flow Modeling. J. Guid. Control. Dyn. 35, 563–577. https://doi.org/10.2514/1.53859.

SESAR Joint Undertaking, 2020. European ATM master plan - Executive view. https://doi.org/10.2829/650097.

Starita, S., Strauss, A.K., Fei, X., Jovanović, R., Ivanov, N., Pavlović, G., Fichert, F., 2020. Air Traffic Control Capacity Planning Under Demand and Capacity Provision Uncertainty. Transp. Sci. 54, 882–896. https://doi.org/10.1287/trsc.2019.0962.

Tandale, M., Menon, P., Rosenberger, J., Subbarao, K., Sengupta, P., Cheng, V., 2008. Queueing Network Models of the National Airspace System, in: The 26th Congress of ICAS and 8th AIAA ATIO. American Institute of Aeronautics and Astronautics, Reston, Virigina, pp. 1–14. https://doi.org/10.2514/6.2008-8942.

Welch, J., 2015. En route sector capacity model final report. ATC-426 URL . https://www.ll.mit.edu/sites/default/files/publication/doc/2018-12/Welch_2015_ATC-8426.pdf (accessed 6.16.20).

Xu, N., Sherry, L., Laskey, K.B., 2008. Multifactor model for predicting delays at U.S. airports. Transp. Res. Rec. J. Transp. Res. Board 2052, 62–71. https://doi.org/10.3141/2052-08.

# Appendices

## A1 Estimated queuing engine parameters of airports and en-route congestion points

Table 7 Estimated parameters of the queuing system at each airport.

| Airport | Service rate (per 15 min) | $k_r$ | Airport | Service rate (per 15 min) | $k_r$ | Airport | Service rate (per 15 min) | $k_r$ |
|---|---|---|---|---|---|---|---|---|
| BAV | 2 | 4 | KHN | 4 | 2 | TAO | 7 | 2 |
| CAN | 14 | 2 | KMG | 13 | 3 | TNA | 6 | 2 |
| CGO | 8 | 2 | KWE | 8 | 3 | TPE | 4 | 3 |
| CGQ | 6 | 3 | KWL | 2 | 2 | TSA | 2 | 3 |
| CKG | 12 | 2 | LHW | 6 | 3 | TSN | 6 | 5 |
| CSX | 8 | 2 | LJG | 3 | 3 | TXN | 2 | 3 |
| CTU | 13 | 3 | LXA | 2 | 5 | TYN | 5 | 2 |
| CZX | 2 | 3 | LYA | 2 | 3 | URC | 8 | 3 |
| DLC | 7 | 3 | LYG | 2 | 3 | WNZ | 4 | 3 |
| DLU | 2 | 4 | MFM | 2 | 3 | WUH | 9 | 3 |
| FOC | 6 | 2 | NGB | 4 | 2 | WUX | 3 | 3 |
| HAK | 8 | 2 | NKG | 9 | 2 | XIY | 12 | 2 |
| HET | 4 | 3 | NNG | 6 | 2 | XMN | 9 | 3 |
| HFE | 4 | 3 | PEK | 18 | 3 | XNN | 3 | 2 |
| HGH | 11 | 2 | SHA | 13 | 2 | XUZ | 2 | 4 |
| HKG | 5 | 3 | SHE | 8 | 3 | YIW | 2 | 4 |
| HRB | 8 | 3 | SJW | 4 | 3 | YNT | 3 | 4 |
| INC | 4 | 2 | SYX | 7 | 2 | ZUH | 4 | 2 |
| JJN | 2 | 4 | SZX | 13 | 2 | | | |

Table 8 Estimated parameters of the queuing system at each en-route congestion point.

| En-route index | Service rate (per 15 min) | $k_r$ | En-route index | Service rate (per 15 min) | $k_r$ | En-route index | Service rate (per 15 min) | $k_r$ |
|---|---|---|---|---|---|---|---|---|
| 1 | 9 | 1 | 11 | 4 | 1 | 21 | 5 | 1 |
| 2 | 20 | 2 | 12 | 7 | 1 | 22 | 3 | 1 |
| 3 | 12 | 1 | 13 | 5 | 1 | 23 | 2 | 1 |
| 4 | 12 | 1 | 14 | 7 | 1 | 24 | 5 | 1 |
| 5 | 7 | 2 | 15 | 5 | 2 | 25 | 3 | 1 |
| 6 | 10 | 1 | 16 | 4 | 1 | 26 | 2 | 1 |
| 7 | 13 | 1 | 17 | 4 | 1 | 27 | 2 | 1 |
| 8 | 11 | 1 | 18 | 8 | 1 | 28 | 4 | 1 |
| 9 | 10 | 1 | 19 | 3 | 1 | 29 | 5 | 2 |
| 10 | 8 | 1 | 20 | 6 | 2 | 30 | 2 | 1 |

## A2 Calculation of flight delay caused by different sources

The total flight delay over a network can be evaluated by whether it is caused by airport or en-route local congestion or network propagation effect, via the following four measures: average local delay at an airport, average propagated delay at an airport, average local delay at an en-route congestion point and average propagated delay at an en-route congestion point. The measures are defined as Eqs. (37) - (40) and the notations are given as Table 5.

- Average local delay at an airport: the average delay incurred on takeoff or landing at an airport.

$$\frac{1}{\|f \in F | d(f) = a\|} \sum_{\{f \in F | d(f) = a\}} W_a(AA(f)). \qquad (40)$$

- Average propagated delay at an airport: the average delay observed at an airport can be attributed to the earlier flights of the same aircraft.

$$\frac{1}{\|f \in F | d(f) = a\|} \sum_{\{f \in F | d(f) = a\}} [AA(f) - SA(f)]. \qquad (41)$$

- Average local delay at an en-route point: the average delay incurred when passing through an en-route congestion point. Note since there is an E_Buffer within each flight (as described in Section 3.2.3), some en-route delay can be absorbed by E_Buffer. Thus, E_Buffer should be subtracted from the effective local delay generated by en-route points.

$$\left\{ \frac{1}{\|f \in F | en(f) = e\|} \sum_{\{f \in F | en(f) = e\}} W_e(AE_e(f)) \right\} - E_{\text{Buffer}}. \qquad (42)$$

- Average propagated delay at an en-route point: the average delay observed at an en-route point can be attributed to the delays incurred by the earlier flight of the same aircraft.

$$\frac{1}{\|f \in F | en(f) = e\|} \sum_{\{f \in F | en(f) = e\}} [AE_e(f) - SE_e(f)]. \qquad (43)$$

Table 9 Notations of network performance measures.

| | |
|---|---|
| $f$ | A flight |
| $F$ | The set of all flights |
| $o(f)$ | Origin airport of $f$ |
| $d(f)$ | Destination airport of $f$ |
| $en(f)$ | An en-route point passed by $f$ |
| $SA(f)$ | Scheduled arrival time of $f$ |
| $AA(f)$ | Adjusted arrival time of $f$, $AA(f) \geq SA(f)$ |
| $SE_e(f)$ | Scheduled arrival time of $f$ at en-route point e |
| $AE_e(f)$ | Adjusted arrival time of $f$ at en-route point e, $AE_e(f) \geq SE_e(f)$ |
| $W_a(t)$ | Waiting time at time $t$ of airport $a$ |
| $W_e(t)$ | Waiting time at time $t$ of en-route congestion point $e$ |
| E_Buffer | En-route buffer time |

## A3 Simulation setup for the stylized network

In order to setup the environment for the simulation method, the flight schedule data is first generated for the stylized network as Fig. 23. Then, a discrete-event simulation is performed using the generated flight schedule data to obtain the delay statistics of the stylized network.

### A3.1 Flight schedule data generation

We let the aircraft departure at A1 follow Poisson distribution with departure rate $\lambda_{dep}$, and the departure time for the aircraft is regarded as the scheduled departure time at A1, noted as $T_{A1}^{dep}$. The aircraft then fly sequentially from A1 to E1, then E1 to A2, and finally A2 to A3. The scheduled times for the airports and en-route congestion point are computed using Eqs. (41) - (44):

$$T_{E1}^{pas} = T_{A1}^{dep} + FT_{A1-E1}, \qquad (44)$$

$$T_{A2}^{arr} = T_{E1}^{pas} + FT_{E1-A2}, \qquad (45)$$

$$T_{A2}^{dep} = T_{A2}^{arr} + STurn_{A2}, \qquad (46)$$

$$T_{A3}^{arr} = T_{A2}^{arr} + FT_{A2-A3,} \qquad (47)$$

where $T_{ser}^{pas}$, $T_{ser}^{dep}$ and $T_{ser}^{arr}$ represent the scheduled passing through time, departure time and arrival time at $ser$, respectively, $ser \in [A1, E1, A2, A3]$. And $FT_{ser1-ser2}$ denotes the average flight time between $ser1$ and $ser2$, $ser1 \in [A1, E1, A2, A3]$, $ser2 \in [A1, E1, A2, A3]$; $STurn_{A2}$ is the schedule turnaround time of A2.

### A3.2 Discrete-event simulation

After the generation of the flight schedule data, the simulation method is used to model the operation of a system as a (discrete) sequence of events in time. The results computed from the simulation model are then used as bench mark results to compare the performance of MATND and AND model. The simulation method is shown as Fig. 36, where the system is first discretized into a list of events ordered by their operation time. Then, the first event of the event list is first performed, and then the system is updated and the times of the following event list are revised, after that, the previous event is marked as processed and the system moves to the next event.

To take into the uncertainty in real operations, we add a Gaussian noise to the actual flight time $\widetilde{FT}_{ser1-ser2}$ between $ser1$ and $ser2$ and actual turnaround time $ATurn_{A2}$, illustrated as Eq. (45) – Eq. (46),

$$\widetilde{FT}_{ser1-ser2} = FT_{ser1-ser2} + \text{Gauss}(\mu_0, \delta_0) \qquad (48)$$

$$ATrun_{A2} = minTurn_{A2} + \text{Gauss}(\mu_0, \delta_0) \qquad (49)$$

where $\text{Gauss}(\mu_0, \delta_0)$ denotes the Gaussian distribution with $\mu_0$ mean and standard deviation $\delta_0$, and $ser1, ser2 \in [A1, E1, A2, A3]$, and $minTurn_{A2}$ represents the minimal turnaround time of A2. Here, $STurn_{A2}$ is set to be 30 min, and $minTurn_{A2} = 25$ min, and $\mu_0 = 0$ and $\delta_0 = 5$.

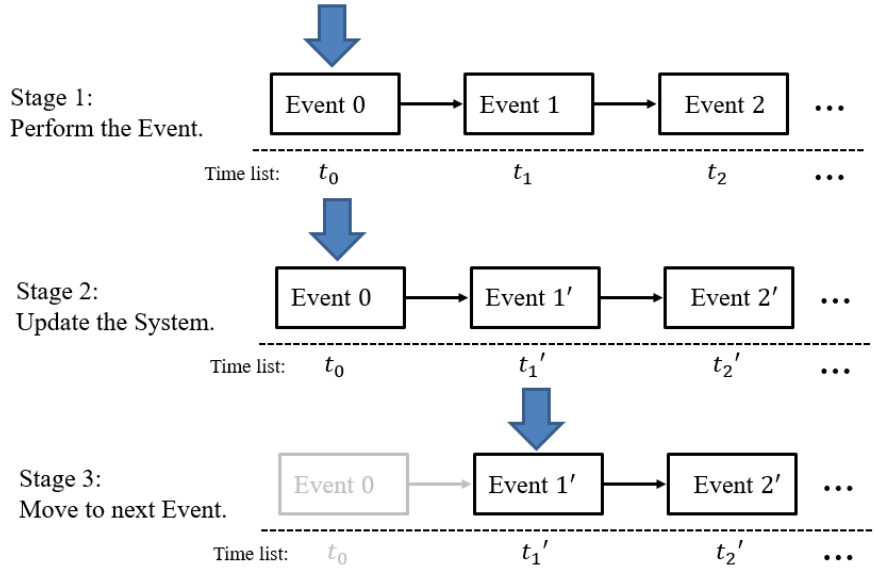

Fig. 36 Demonstration of the simulation procedures.

The notation of event and the method to update the system of the stylized network are described as below. Note, we use the term "request the service of an airport/en-route congestion point" when an aircraft departs/arrives at an airport/en-route congestion point.

1. Event

    a. The process that an aircraft requests the service of an airport/en-route congestion point.

    b. The process that an airport/en-route congestion point finishes its service for an aircraft.

2. Event time

    a. The actual departure/arrival time of the aircraft.

    b. The time of the airport/en-route congestion point finishing its service for an aircraft.

3. The method to update the system

    a. Queue engine at each airport / en-route congestion point.

## A4 Throughput and delay predictions at CAN

The throughput and delay predictions at CAN in November, 2016 using MATND and AND methods, compared to the actual data are provided below to give a more comprehensive view of the model performance.

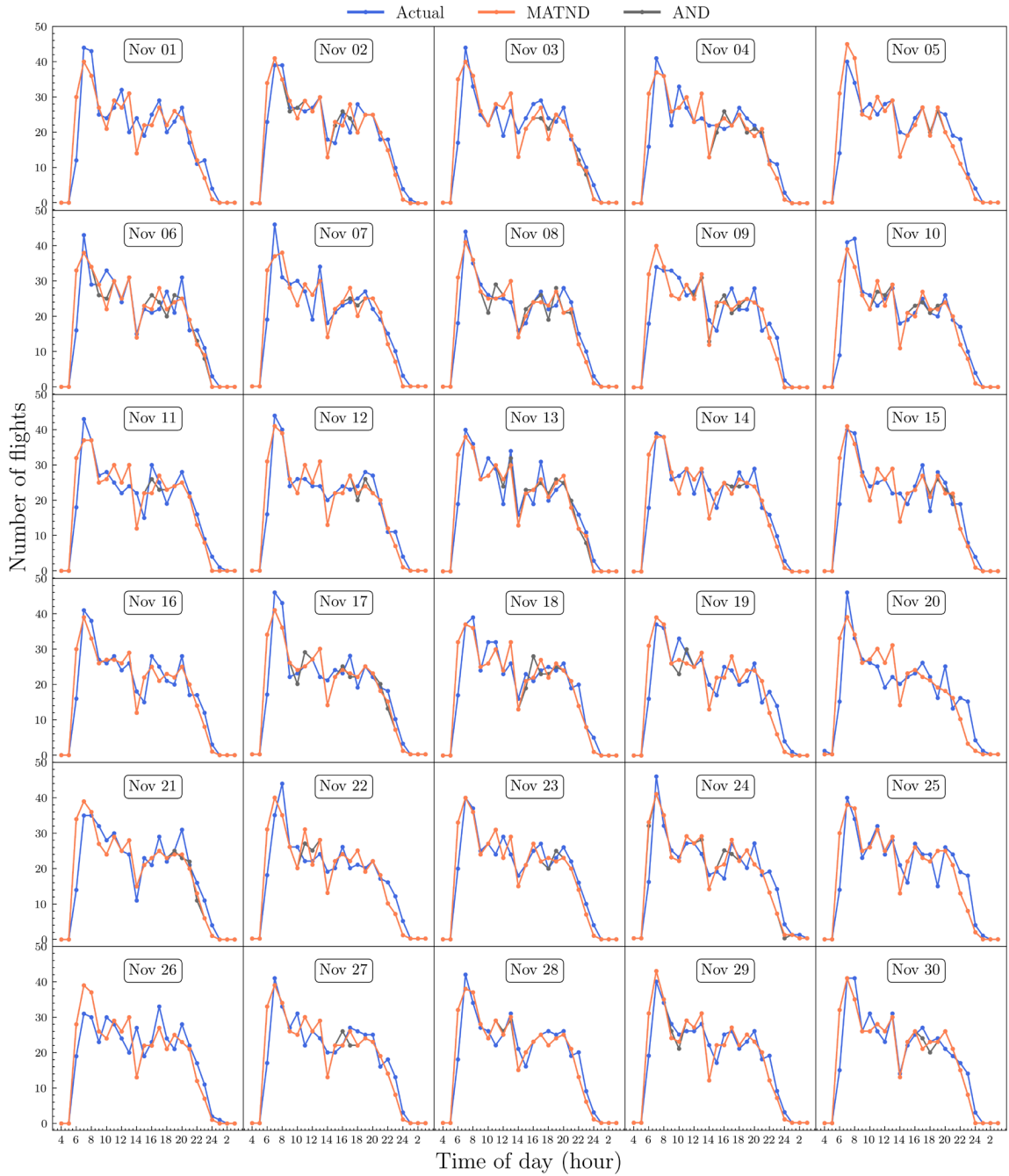

Fig. 37 Departure throughput at CAN in Nov, 2016.

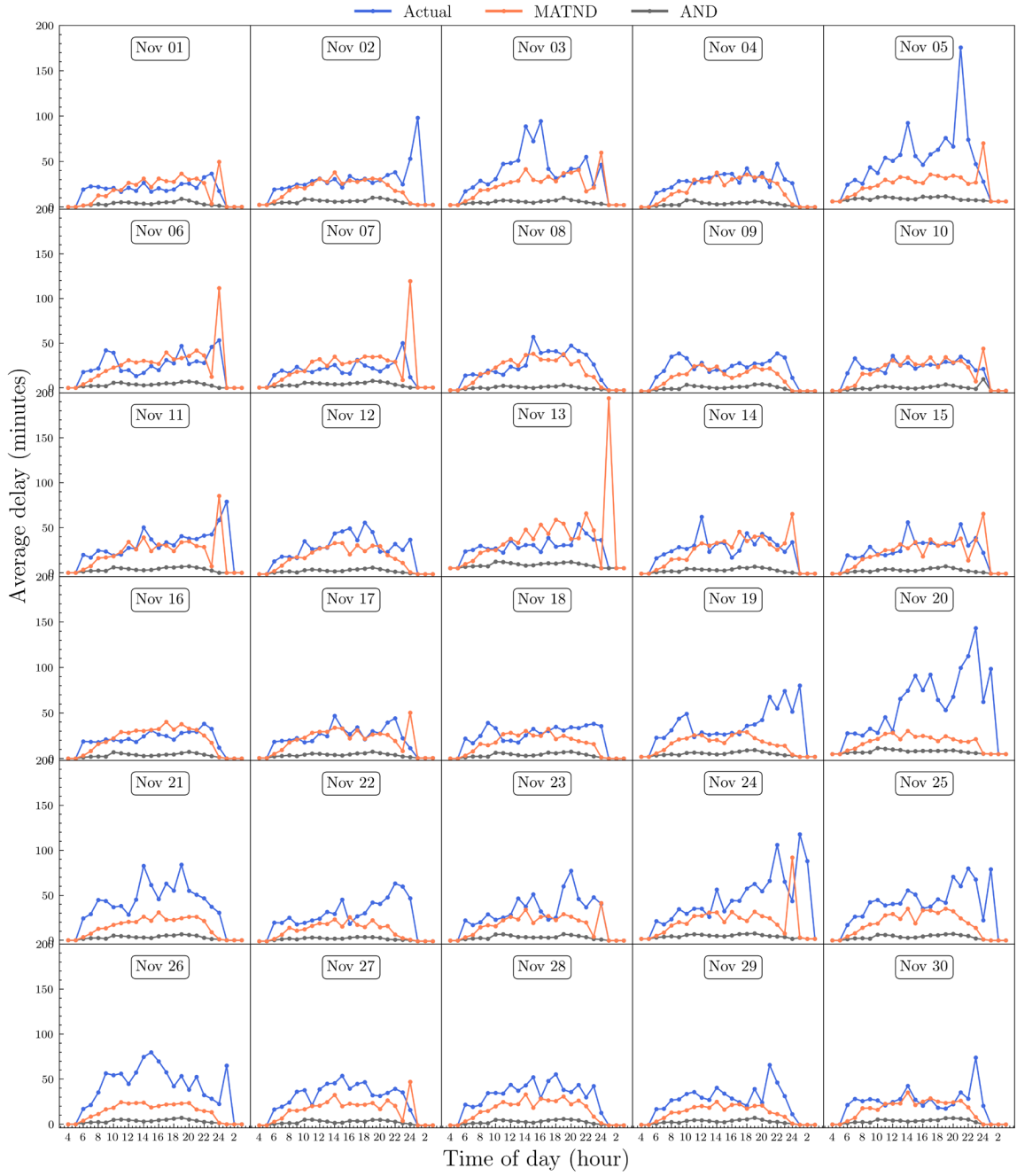

Fig. 38 Departure delay at CAN in Nov, 2016.

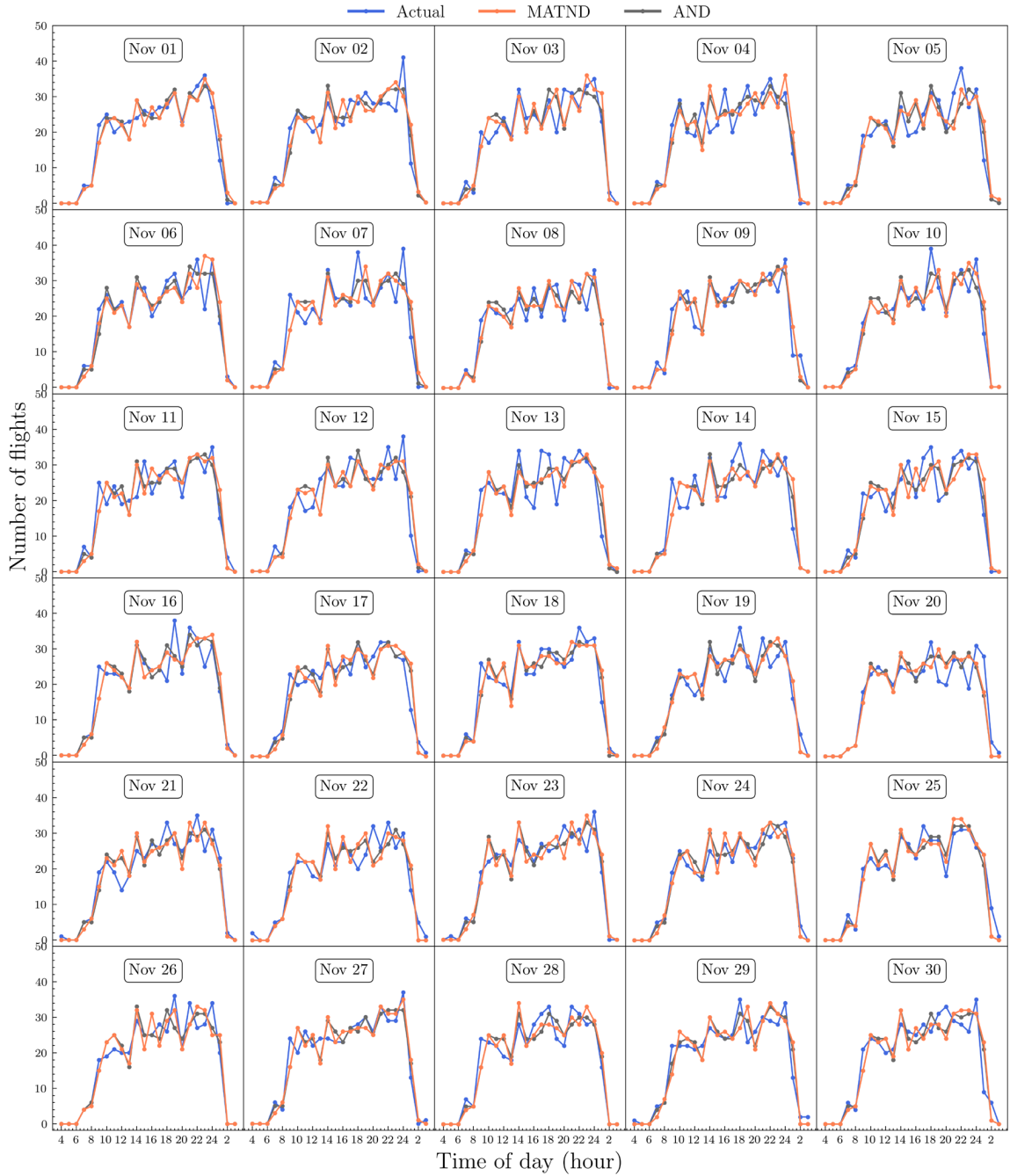

Fig. 39 Arrival throughput at CAN in Nov, 2016.

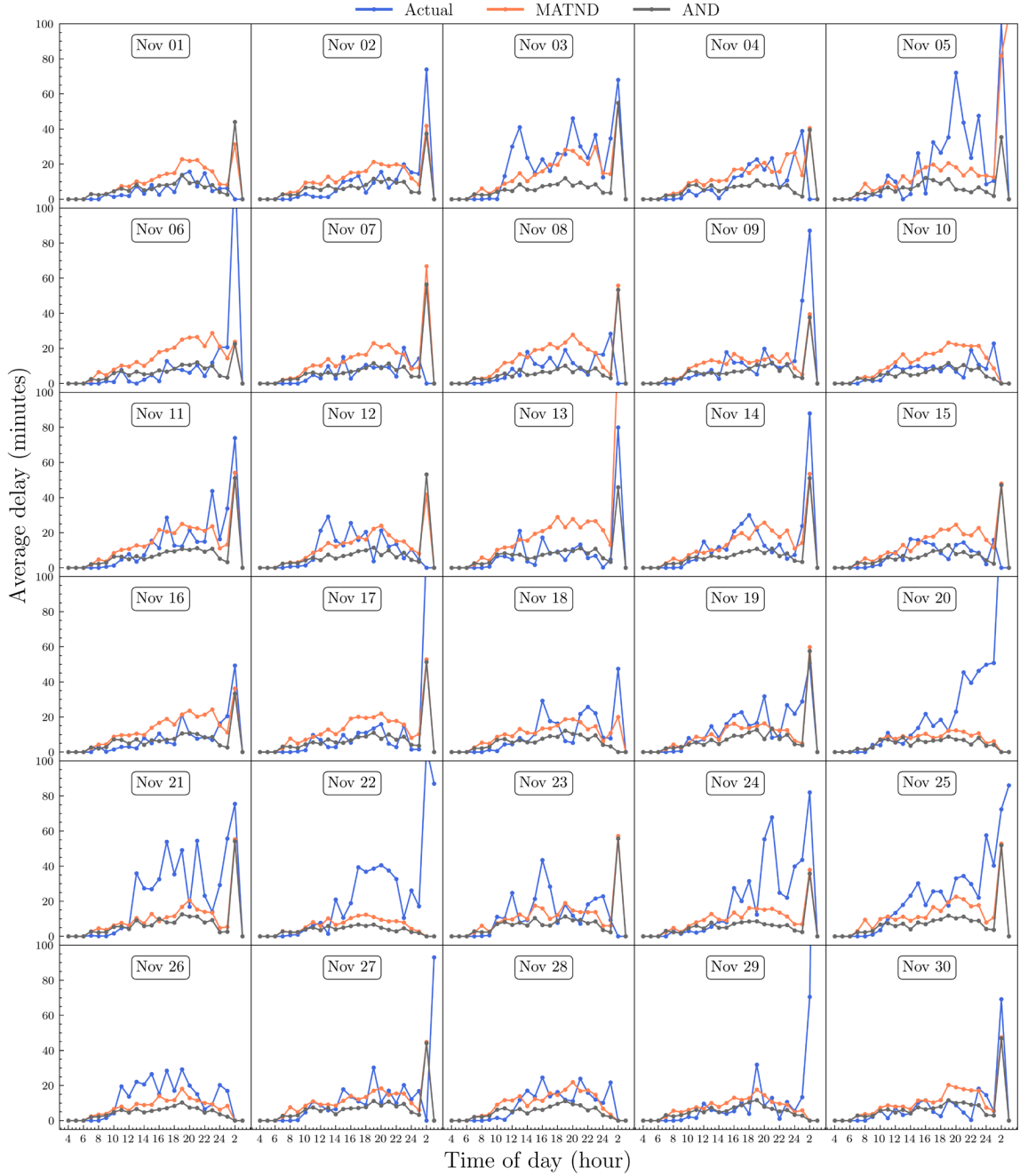

Fig. 40 Arrival delay at CAN in Nov, 2016.